\documentclass[reprint, amsmath,amssymb,aps]{revtex4-2}

\usepackage{graphicx}
\usepackage{dcolumn}
\usepackage{bm}
\usepackage{amsmath}
\usepackage[usenames,dvipsnames]{xcolor}
\usepackage{mathtools}
\usepackage[colorlinks=true,citecolor=blue,linkcolor=blue]{hyperref}
\usepackage[caption=false]{subfig}
\usepackage{bbm}
\usepackage{comment}
\usepackage{hyperref}
\usepackage{upgreek}
\usepackage{esint}
\usepackage{soul}
\usepackage[normalem]{ ulem } 
\usepackage{floatrow}
\usepackage{verbatim}
\usepackage{braket}

\usepackage{array}
\newcommand{\PreserveBackslash}[1]{\let\temp=\\#1\let\\=\temp}
\newcolumntype{C}[1]{>{\PreserveBackslash\centering}p{#1}}

\newcommand{\rom}[1]{\uppercase\expandafter{\romannumeral #1\relax}}

\begin{document}

\title{Scalable architecture for trapped-ion quantum computing using RF traps and dynamic optical potentials}

\author{David Schwerdt$^{1,\ast,\dagger}$}
\author{Lee Peleg$^{1,\ast}$}
\author{Yotam Shapira$^{1,2,\ast}$}
\author{Nadav Priel$^{2,\ast}$}
\author{Yanay Florshaim$^{2}$}
\author{Avram Gross$^{2}$}
\author{Ayelet Zalic$^{2}$}
\author{Gadi Afek$^{2}$}
\author{Nitzan Akerman$^{1}$}
\author{Ady Stern$^{3}$}
\author{Amit Ben Kish$^{2}$}
\author{Roee Ozeri$^{1,2}$}

\affiliation{\small{$^1$Department of Physics of Complex Systems, Weizmann Institute of Science, Rehovot 7610001, Israel\\
$^2$ Quantum Art, Ness Ziona 7403682, Israel\\
$^3$ Department of Condensed Matter Physics, Weizmann Institute of Science, Rehovot 7610001, Israel
\\ 
$^*$ These authors contributed equally to this work\\
$^\dagger$ Corresponding author
}}

\begin{abstract}
Qubits based on ions trapped in linear radio-frequency traps form a successful platform for quantum computing, due to their high fidelity of operations, all-to-all connectivity and degree of local control. In principle there is no fundamental limit to the number of ion-based qubits that can be confined in a single 1D register. However, in practice there are two main issues associated with long trapped-ion crystals, that stem from the ‘softening’ of their modes of motion, upon scaling up: high heating rates of the ions' motion, and a dense motional spectrum; both impede the performance of high-fidelity qubit operations. Here we propose a holistic, scalable architecture for quantum computing with large ion-crystals that overcomes these issues. Our method relies on dynamically-operated optical potentials, that instantaneously segment the ion-crystal into cells of a manageable size. We show that these cells behave as nearly independent quantum registers, allowing for parallel entangling gates on all cells. The ability to reconfigure the optical potentials guarantees connectivity across the full ion-crystal, and also enables efficient mid-circuit measurements. We study the implementation of large-scale parallel multi-qubit entangling gates that operate simultaneously on all cells, and present a protocol to compensate for crosstalk errors, enabling full-scale usage of an extensively large register. We illustrate that this architecture is advantageous both for fault-tolerant digital quantum computation and for analog quantum simulations.

\end{abstract}

\maketitle

\section{Introduction}

Trapped ions have ideal properties to be used as qubits for quantum computing (QC); they feature long coherence times, efficient state preparation and detection techniques, and a high degree of connectivity \cite{gaebler2016high,ballance2016high,wang2021single,clark2021high,srinivas2021high}. A quantum register of 1000s of qubits, or more, can be formed, for example, by utilizing an equally spaced crystal of ions in a linear RF Paul trap. Indeed recent years have seen many experimental attempts to work with increasingly larger trapped-ion registers \cite{pagano2019quantum, pogorelov2021compact, yao2022experimental,kranzl2022controlling,guo2023siteresolved}. \par

However, there are two practical issues associated with large ion-crystals that impede progress in this direction. The first is heating rates; as the number of ions in the crystal, $N$, increases, heating of the ions’ motional modes due to electric field noise drastically increases. Of particular concern is the axial center-of-mass (COM) mode, whose frequency typically decreases as $1/N$. This mode is especially vulnerable, as electric field noise tends to be spatially uniform and to target low-frequency modes \cite{turchette2000heating,brownnutt2015ion}. The resulting significant heating rates prohibit the implementation of high fidelity qubit operations, and might destabilize the ion-crystal. \par

The second issue with large ion-crystals is spectral crowding. As the size of the crystal increases, the frequency of adjacent motional modes becomes tightly spaced. For large ion-crystals with a dense mode spectrum, resolving individual modes becomes challenging; this complicates the implementation of entangling gates between two or more ions, which involves exciting their common motion. Granted, spectral control methods \cite{shapira2020theory,shapira2023programmable,shapira2023fast} allow for simultaneously targeting multiple modes to achieve desired qubit couplings with high fidelity. While such methods are promising for a moderate number of ions (up to 100s), they do not generally provide a scalable solution for arbitrarily large ion-crystals. First, the optimal control problems that must be solved to implement these methods becomes intractable for large $N$. Moreover, there is strong evidence that the minimum achievable gate time is set by the smallest frequency spacing among the motional modes \cite{shapira2023fast}. This implies that gate times scale at least as $N^2$, making large ion-crystals prohibitively slow for quantum computations. \par

One active direction for scaling ion traps is the quantum charge-coupled device (QCCD) architecture \cite{kielpinski2002architecture,moses2023race}. This setup involves many spatially separated trapping sites, each containing a small number of ions, where communication between sites is done by shuttling individual ions. Inevitably, this comes at the expense of high overhead in hardware and long circuit duration dominated by ion-shuttling and ion-cooling times. An alternative scale-up approach is using photonic-interconnects in order to link small-scale ion-crystals \cite{monroe2014photons,nadlinger2022experimental,krutyanskiy2023entanglement}. This method likewise involves a high overhead, due to the currently low entanglement rate via the interconnect, leading to slow operations.\par

Here we propose a scalable architecture for QC based on trapped-ion qubits that maintains the advantages of a long ion-crystal while circumventing its challenges. In our proposal, an arbitrarily long ion-crystal is segmented into re-configurable cells by means of dynamically operated optical potentials, e.g. optical tweezers. In this way, the crystal’s motional mode structure is modified such that heating rates reflect only the cell size, and not $N$. Furthermore, programmable high-fidelity multi-qubit entangling gates can be implemented independently within each cell simultaneously. In addition, we show that mid-circuit measurements, a cornerstone of quantum error correction (QEC) techniques, as well as other central quantum computational tools, are straightforward to implement in an optically segmented ion-crystal. \par

In this work we generalize the method presented in Ref. \cite{shapira2023fast}, and employ qubit-local driving fields that implement programmable multi-qubit entangling gates. The method in Ref. \cite{shapira2023fast} provides a clear physical intuition, an efficient procedure for designing gate drive parameters, and a thorough scaling analysis - all of which are relevant to this proposal.  \par

The integration of optical tweezers into ion traps is an active area of research \cite{olsacher2020scalable,shen2020scalable,espinoza2021engineering,mazzanti2021trapped,teoh2021manipulating,mazzanti2023trapped}. Specifically Refs. \cite{espinoza2021engineering,mazzanti2021trapped,teoh2021manipulating,mazzanti2023trapped} make use of optical tweezers in order to generate target entanglement operations. In Ref. \cite{olsacher2020scalable} the authors use tweezers for parallel two-qubit entangling gates, the system's scalability is discussed however there is no treatment of large ion segments or ion-crystal heating rates. In Ref. \cite{shen2020scalable} the authors consider segmenting an ion-crystal using tweezers, however do not analyze generating large-scale entanglement or connectivity between the segments. Mid-circuit measurements are not discussed in any of the above works. Unlike previous studies, here we detail how to achieve a holistic architecture, incorporating parallel and large-scale  entangling operations and mid-circuit measurements, that are needed for large scale QC. \par

The remainder of this paper is organized as follows. In section II we present an overview of the proposed architecture. In section III we sketch two examples of applications of quantum information processing tasks that are amenable to our architecture: quantum simulation and quantum error correction. An in-depth analysis of our proposed architecture follows. Namely, in section IV we derive the spectral properties of optically segmented traps. In section V we analyze the implication of optical segmentation on the ion-crystal's heating rate. In section VI we present our method for designing high-fidelity multi-qubit logical operations that are not hindered by unwanted crosstalk. Lastly, in section VII we present a protocol for performing mid-circuit measurements. \par

\section{Proposed architecture}
There are well-established techniques for preparation, control, and measurement of trapped-ion based qubits for small-scale registers \cite{ozeri2011trapped,bruzewicz2019trapped}. We therefore focus mostly on unique aspects of our architecture. We propose a QC architecture in which a long trapped-ion crystal, made of $N$ ions, is segmented into $S$ cells, with each cell containing $C$ computational qubits available for quantum computation and simulation, generating a $C\cdot S$ qubit register. The segments are formed by placing $B_A$ barrier ions between adjacent cells, that are illuminated by an optical trapping potential, i.e. optical tweezers, made of tightly focused laser beams \cite{ashkin1970acceleration,chu1985three}, that provide an additional local confinement for the barrier ions. \par

The key point enabling scalability is that due to the optical segmentation the cells behave as nearly independent quantum registers. Indeed, we show below that both challenges of scaling-up trapped-ion based QCs, i.e., ion-crystal stability and the performance of logical operations, both scale with the cell size, $C$, and not the total number of cells, $S$. Thus the dynamically reconfigurable segmentation removes the fundamental limits on the number of cells and on the number of qubits in the ion-crystal, and accordingly on quantum circuit and simulation size.\par

\begin{figure}[!hbtp]
	\caption{Architecture for scalable trapped-ion quantum computing using RF traps and dynamic optical potentials. The ion-crystal (vertical direction) is segmented into cells by dynamically applying optical trapping potentials. Sequential applications of parallel quantum operations and reconfiguration of the optical trapping implements a large-scale quantum circuit on the entire ion-crystal. We highlight a section of the quantum circuit and the sequence used. As shown, optical trapping generates $S$ segments, each containing $C$ computational qubits (white circles) that are separated by $B_A$ consecutive optically confined ‘barrier ions’ (purple filled circles). Additional $B_B$ ions in each cell (red circles) are used to support additional segmentation configurations of the ion-crystal, and mid-circuit measurements that enable classical feedback (red lines) for the implementation of e.g. quantum error correction. The horizontal direction shows a typical mode of operation of the QC, namely logical operations, involving multi-qubit and single-qubit logical gates, $U_{s,n}^{\left(A\right)}$, are performed in each cell, $s$, at the $n$th computational step. The optical potentials are then switched to a different segmentation configuration, allowing for large-scale connectivity between cells of the previous configuration,  $U_{s,n}^{\left(B\right)}$. An intermediate segmentation configuration, in which all $B_A$ and $B_B$ ions are optically confined, is used to accommodate for mid-circuit measurement and state re-initialization. Additional computational steps ($n+1$, etc.) can be performed as required, thus ultimately guaranteeing connectivity between all computational qubits. }
	\centering
	\includegraphics[width=1\linewidth]{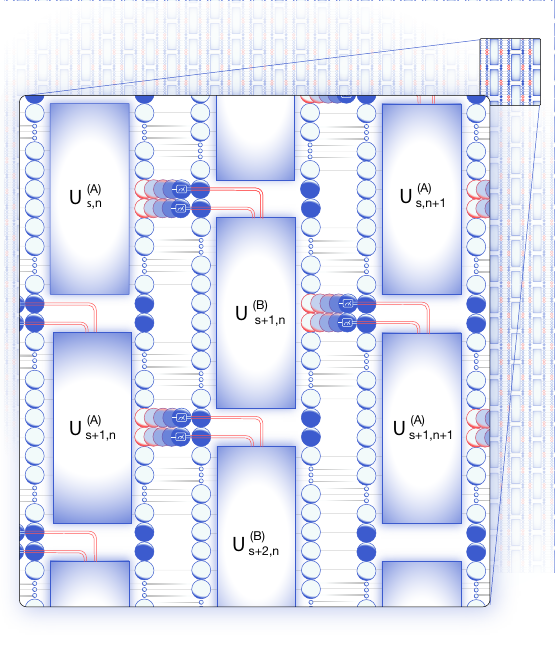}
	\label{figMain}
\end{figure}

A general overview of our scalable architecture, is presented schematically in Fig. \ref{figMain}. In favor of a simple presentation we focus on a section of the entire ion-crystal, and a section of the operations used. We explicitly highlight $C=8$ computational qubits (white circles) per cell. In practice 10s of qubits may be placed in each cell, and the number of cells is theoretically unlimited (vertical dots), forming a large ion-crystal. The cells are separated by $B_A$ barrier ions (purple filled circles), with the schematic showing $B_A=2$. $B_B$ additional ions (red circles) are placed in the center of each cell and are used for segment re-configuration and mid-circuit measurements, with the schematic showing $B_B=2$. Thus, the ion-crystal contains $N=\left(C+B_A+B_B\right)S+B_A$ trapped ions. \par

The figure shows a general mode of operation of the QC, with parallel operations along the vertical axis and sequential operations, ordered from left to right, that form computational steps. Specifically, barrier ions are strongly confined with optical trapping potentials (purple filled circles), generating the $S=2$ segments in this example (configuration ‘A’). At the $n$th computational step, local and global driving fields  simultaneously implement unitary operations, $U_{s,n}^{(A)}$, with $s=1,2,...,S$. These unitary operators implement components of the overall quantum algorithm, and may involve all $C$ computational qubits (white circles) and $B_B$ auxiliary ions  (red circles) which are not optically confined in this configuration. \par

Next all $B_A$ and $B_B$ barrier ions are optically confined, enabling mid-circuit measurement of the auxiliary ions as well as preparation of ions to be used as auxiliary ions (detailed below). Classical feedback (red lines), based on the mid-circuit measurement results, may be implemented at this step in order to influence the next layer of unitary evolution. \par

Next, the segmentation is reconfigured by removing the confinement of the $B_A$ barrier ions. This dynamically and quickly generates a different configuration of the segments (configuration ‘B’). Unitary operators, $U_{s,n}^{(B)}$, can now be implemented and connect previously uncoupled qubits. Additional measurement and reconfiguration steps can be applied as required. We remark that here we have shown only two basic ion-crystal configurations (A and B), and have further considered here and below $B_A=B_B$, however various additional segmentation configurations, with different parameters (e.g. segment and barrier sizes), can be flexibly generated provided relevant ions are allocated as barrier ions. \par

In addition to the configurable optical trapping potentials, we make use of local independently-applied fields for driving programmable multi-qubit gates \cite{shapira2023fast} ($U^{(A)}$s and $U^{(B)}$s in Fig. \ref{figMain}), as well as for mid-circuit measurements and state preparation.\par

\section{Application demonstrations}
We sketch examples of applications that showcase the utility of our architecture. These examples rely on various features of our proposal, e.g. implementing large-scale multi-qubit entangling operations, register reconfiguration, and performing mid-circuit measurements. The methods underlying these features are thoroughly discussed and analyzed in the sections below. \par

We start by considering analog quantum simulations on our system, specifically a three dimensional (3D) quantum-spin model. In general, quantum simulations are considered a well-suited task for noisy intermediate-scale quantum (NISQ) era QCs \cite{preskill2018quantum}, as simulation of quantum systems is a notoriously challenging task for classical computers, while quantum computers are considered naturally suited for it \cite{feynman1982simulating}. Indeed, numerous quantum systems have recently implemented impressive demonstrations of quantum simulations \cite{neill2021accurately,semeghini2021probing,kim2023evidence}. \par

Quantum simulations using trapped-ion based qubits, have also been recently demonstrated \cite{kokail2019self,tan2021domain,kyprianidis2021observation,joshi2022observing,
qiao2022observing,shapira2023quantum,iqbal2023creation,wu2023qubits,schuckert20231dsim}, many of which take advantage of the unique all-to-all coupling present in these systems. In these works Ising-type interactions, that are inherent to trapped-ion systems, can be straightforwardly iterated and combined with single-qubit rotations in order to generate arbitrary XYZ-type spin-models \cite{porras2004effective,arrazola2016digital}. \par

Our architecture is well-suited for quantum simulations, as our ability to design programmable multi-qubit entanglement manifests as the dimensionality and geometry of the simulated model \cite{manovitz2020quantum}, and the optical segmentation and reconfiguration provide a straightforward approach to simulating 3D systems. Such 3D systems are challenging to simulate on linear or planar quantum processors with short-range interactions, due to a large gate overhead required for embedding the 3D geometry. \par

\begin{figure}[!hbtp]
	\caption{Quantum simulation of a 3D rectangular Ising spin-model on an optically segmented ion-crystal. Computational qubits (white circles) are coupled in cells (purple brackets) such that each cell holds two planes of the rectangular spin-model, and implements inter-plane (dark-blue and bright-blue links) and intra-plane (red links) couplings (a) Couplings of a single cell, holding two planes of the model, with $d=4$, used for both segmentation configurations. (b) Multiple cells form the 3D Ising model. The two optical configurations (‘A’, ‘B’ and purple brackets) enable interlaced coupling of the planes (solid red and dashed red, respectively), such that two entanglement operations implement the model's Hamiltonian. (c) We design the control pulses that implement the nearest-neighbor couplings required for the Ising model, here for $d=4$ and $S=5$. The resulting coupling between all qubit pairs (horizontal and vertical axes) is shown (color), exhibiting a block-diagonal structure that implements the links shown in (a), for each cell. (d) Deviation of the qubit-qubit coupling from the ideal structure (log scale), showing a low error, that is not limited by crosstalk between adjacent cells, and enables scaling up. The implementation's infidelity scales as the deviations squared, and is here evaluated to $10^{-4}$ per cell.} 
	\centering
	\includegraphics[width=1\linewidth]{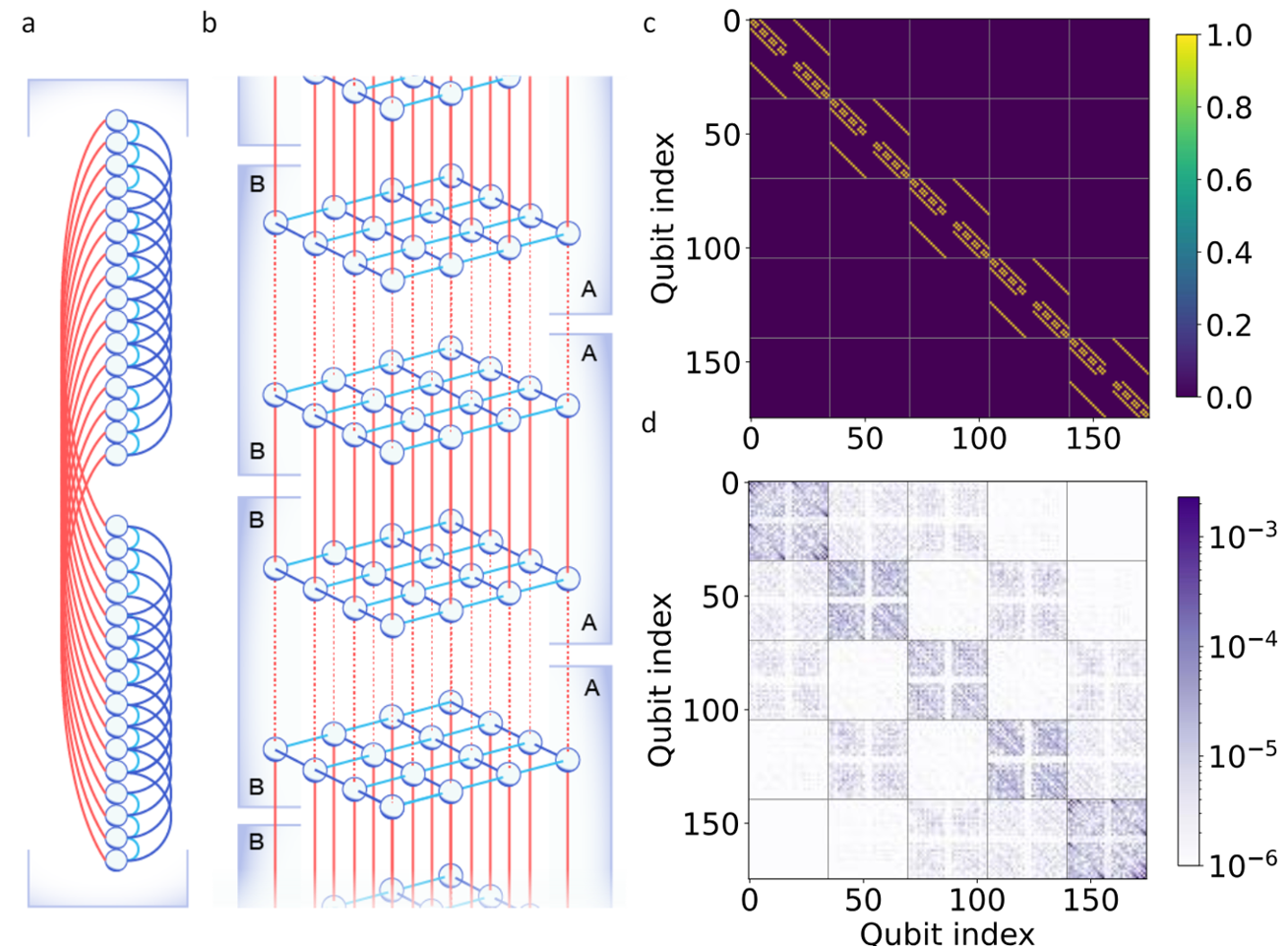}
	\label{figSimulation}
\end{figure}

For simplicity, we focus here on a 3D rectangular lattice of spins, though many other geometries can be realized, e.g. hexagonal. Figure \ref{figSimulation} shows an implementation of such a $d\times d\times2S$ rectangular Ising model. Specifically, Fig. \ref{figSimulation}(a) shows a single cell of the ion-crystal, that contains $2d^2$ computational qubits (white circles), and encodes a pair of two-dimensional planes of the simulated system. Programmable entanglement gates implement the required couplings that generate the model's geometry, i.e. inter-plane couplings (dark-blue and bright-blue) and intra-plane couplings (red). The model is made three dimensional, as shown in Fig. \ref{figSimulation}(b), by interlacing  the vertical couplings between adjacent layers (solid red and dashed red), using the two segmentation configurations (purple brackets). \par

As an example, we consider a specific realization of our architecture, namely qubits implemented on the $4S_\frac{1}{2}$ Zeeman ground-state manifold of trapped $^{40}\text{Ca}^+$ ions, coupled by Raman transitions mediated by the $4P$ manifolds. We utilize this realization for a $4\times4\times10$ nearest-neighbor Ising model, i.e. $d=4$ and $S=5$ (see further system details in the SM \cite{SM}). We design the control pulses that drive the computational qubits, and generate the required entanglement operations for simultaneous $XX$-coupling of the qubits, in the entire ion-crystal, similarly to the couplings shown in Fig. \ref{figSimulation}(a). For simplicity, we use a uniform, unity coupling between all nearest-neighbor qubits. Our design protocol is presented and discussed below. \par

We calculate the coupling terms that result from the control pulses that we designed, and plot them in Fig. \ref{figSimulation}(c), with each point on the plot representing a qubit-qubit coupling between qubits of the corresponding indices (horizontal and vertical axes). The block-diagonal structure that is seen reflects the underlying segmented structure of the ion-crystal. The relative difference between the ideal and the resulting qubit-qubit couplings is shown in Fig. \ref{figSimulation}(d) in log-scale exhibiting a low coupling error that indicates a high-fidelity and accurate simulation. Indeed, the overall performance of our implementation, simultaneously coupling $160$ computational qubits, is evaluated with an infidelity that is better (lower) than $10^{-4}$ per cell. \par

We note that with this encoding, a time-step of an $XX$ nearest-neighbor 3D rectangular Ising model can be implemented with only two sequential multi-qubit entanglement gates, such as that shown in Fig. \ref{figSimulation}(c), which, for $d=4$ and $S=5$, would otherwise require the application of 384 sequential two-qubit gates between arbitrary pairs in the quantum register (in general $\mathcal{O}\left(d^2 S\right)$ two-qubit gates). By considering more elaborate models, e.g. next-nearest neighbor interactions within the two-dimensional planes, non-uniform couplings between the planes, or larger systems, the two-qubit gate count will increase, while with our method two multi-qubit gates still suffice. \par

\begin{figure}[!hbtp]
	\caption{Quantum error correction code on an optically segmented ion-crystal. (a) We implement several distance-five logical qubits, shown pictorially as grids of $5\times5$ qubits (white circles). Each cell in configuration ‘A’ (purple brackets, 'A') houses a single logical qubit (labeled 0 - 12 and 25 - 36) and twelve auxiliary qubits (labeled 13 - 24). Programmable multi-qubit gates generate the required entanglement structure for stabilizer measurements on each logical qubit in parallel. Here we show an implementation that generates $X$-stabilizers (light-green plaquettes and blue lines). $Z$-stabilizers may be implemented with an additional gate (light-red plaquettes). Connectivity across the entire register is enabled by configuration ‘B’ (purple brackets, 'B') such that each cell in this configuration holds data qubits of two adjacent logical qubits. This enables logical entangling operations between any neighboring encoded qubits via stabilizer measurements formed on the border between the two surface codes (red plaquettes and blue lines), using the auxiliary ions in this configuration (labeled 17'-19'). (b) Example of programmable multi-qubit gate generating $X$ and $Z$ stabilizers on five logical qubits, implemented on a segmented ion-crystal with $S=5$, $C=37$ and $B=3$. We design the control pulses that implement the entanglement gates required for the various stabilizers. The resulting qubit-qubit couplings between corresponding qubit pairs (horizontal and vertical axes) is shown (color), exhibiting a block-diagonal structure that reflects the formation of five logical qubits. (c) Deviation of the qubit-qubit coupling from the ideal structure (log scale), showing a low error, that is not limited by crosstalk between adjacent cells, and enables scaling up. The gate's infidelity scales as the deviations squared, and is here evaluated to $10^{-4}$ per cell.} 
	\centering
	\includegraphics[width=1\linewidth]{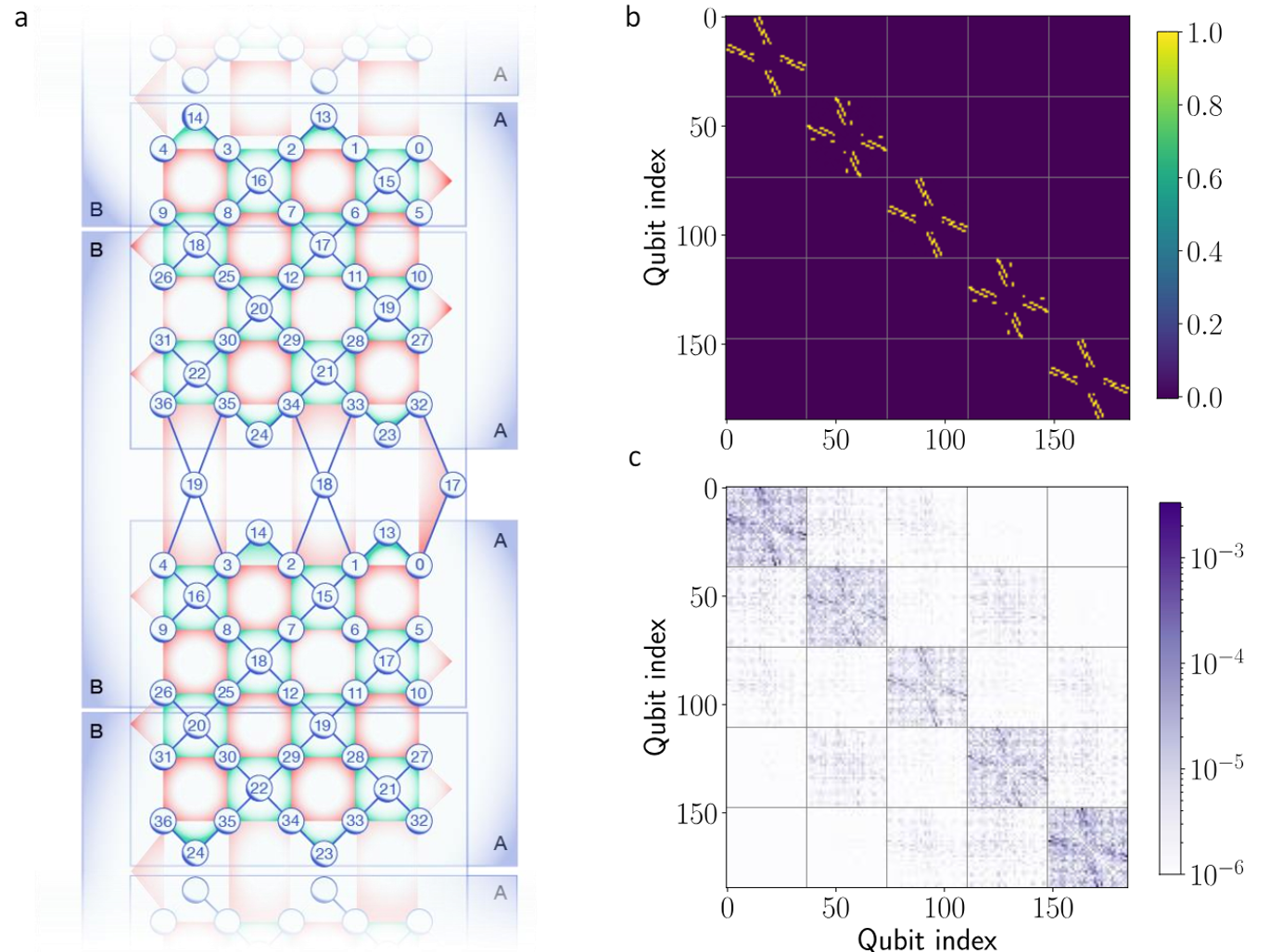}
	\label{figQEC}
\end{figure}

The second example showcases a path towards fault-tolerant quantum computation \cite{knill1998resilient,kitaev2003fault,aharonov2008fault}. As shown in Fig. \ref{figQEC}(a), each segment (purple brackets, ‘A’) is utilized to encode a single logical error-corrected qubit using a distance-five surface code. The code uses 25 computational qubits to store the logical state, and 12 auxiliary (ancilla) qubits to readout the values of both $X$ and $Z$ stabilizer measurements. The entire set of $X$ stabilizer measurements can be implemented in one step, using a single multi-qubit entanglement gate and mid-circuit measurements. Then, the auxiliary qubits are reinitialized, after which the entire set of $Z$ stabilizers is similarly performed. Clearly cell segmentation does not require 12 barrier ions, thus here most of the 12 auxiliary qubits are implemented by computational qubits in the cell. \par

Naturally, these stabilizer measurements can be performed across all logical qubits in parallel. Note that not all segments must implement the same operation at any given time; some may, for example, implement $X$ stabilizer measurements while others implement $Z$ stabilizer measurements. Figure \ref{figQEC}(b) shows a specific example involving five logical qubits, and multi-qubit entanglement gates, designed to perform simultaneous $X$ and $Z$ stabilizer couplings, in different cells, throughout the entire ion-crystal. The relative difference between the ideal and the resulting qubit-qubit couplings is shown in Fig. \ref{figQEC}(c) in log-scale, exhibiting a low error that indicates a low infidelity that is better (lower) than $10^{-4}$ per cell. \par

Entanglement between two logical qubits, implemented on independent surface codes, can be achieved, for example, using  ‘lattice surgery’ \cite{horsman2012surface,erhard2021entangling}. In this method, stabilizers formed along the border between two surface codes are measured, projecting the joint state of the logical qubits to an entangled basis. In our case, this is naturally enabled by re-configuring the optical segmentation (purple brackets, ‘B’) such that ions from neighboring cells can interact, as shown schematically in Fig. \ref{figQEC} (blue lines and red plaquettes in the center of the 'B' configuration). Following this operation, the segmentation reverts to its original configuration, and additional stabilizers are measured. The overall result of this procedure is an entangling logical $XX$ operation between the adjacent surface codes. \par

We emphasize that even though five cells are used in this example, our architecture's scalability enables using considerably more cells in order to implement many logical qubits. This example shows one potential route to fault tolerance using our architecture; however many different variants could be considered. First the distance of the code, and thus the number of physical qubits required, should be determined by the demands on the overall logical error rate. Moreover, it is possible to encode two (or more) logical qubits per segment; this would then enable performing logical entanglement operations transversally, as an alternative to lattice surgery. Finally, the choice of the surface code itself, while motivated by its high fault tolerance threshold, is not the only possible choice. In fact, there are other high-threshold codes, with potentially better encoding rates, that may be suitable as well \cite{bravyi2023highthreshold}. Constructing an optimal protocol for fault tolerant QC within our architecture, including a prescription for performing logical non-Clifford operations, is a subject for further study. \par

An in-depth numerical analysis of the QEC circuit can be found in the SM \cite{SM}. The analysis focuses on  technical aspects of the circuit implementation; it accounts for various heating and cooling mechanisms and provides an estimated resource cost associated with each step in the circuit, and shows that our architecture is experimentally feasible with current state-of-the-art hardware. Furthermore, in our analysis we contrast our architecture with QCCD and photonic-interconnect architectures, showing that our proposal offers up to two orders of magnitude of speed up in implementation of QEC rounds, in terms of overall circuit duration.  

\section{Spectral properties of optically-segmented ion-crystals}
In general, similar to alternative scale-up methods, segmentation decouples cells from each other, reshaping the well-known all-to-all coupling of trapped-ion crystals to local all-to-all couplings within cells of manageable sizes of tens of ions. As we show below, this segmentation preserves the ion-crystal's stability and enables programmable multi-qubit quantum gates that act simultaneously and independently within the different cells. Another unique feature of our scale-up strategy, compared to other segmentation approaches, stems from the ability to dynamically and quickly reconfigure the applied segmentation, generating new cells that combine and couple previously decoupled qubits, thus enabling fast and large-scale connectivity within the whole ion-crystal. \par

We analyze the effect of optical segmentation on the ion-crystal, specifically on collective modes of motion of the ion-crystal in the axial and transverse directions. To this end we assume that the barrier ions are illuminated by beams that generate an optical trapping potential (o.t.p), inducing the trapping frequency \cite{grimm2000optical},
\begin{equation}
	\omega_{\text{o.t.p}} = \sqrt{\frac{2\text{Re}[\alpha\left(\lambda\right)]}{m}}|E|,\label{eqOmegaOtp}
\end{equation}
where $\alpha\left(\lambda\right)$ is the wavelength-dependent polarizibility of an illuminated ion, $m$ is its mass, and $E$ is the field strength. The impact of this potential is best quantified by comparing it to another important parameter in the system, namely the characteristic frequency-scale associated with the Coulomb interaction between adjacent ions \cite{olsacher2020scalable}, 
\begin{equation}
	\nu = \sqrt{\frac{e^2}{4\pi\epsilon_0md^3}},\label{eqNu}
\end{equation}
with $e$ the elementary electron charge, $\epsilon_0$ the vacuum permittivity and $d$ is the inter-ion distance of an equidistant ion-crystal. As we show below, a strong optical potential, $\omega_\text{o.t.p}>\nu$, drastically changes the motional mode structure of the chain. By incorporating the optical trapping potential with common methods to analyze ion-crystal motion \cite{johanning2016isospaced}, we obtain the normal-mode frequencies and structure in the axial and transverse directions (for further details see the SM \cite{SM}). In the following, we assume that that the optical trapping potential is equally strong in the axial and a single transverse direction, and zero in the other transverse direction. In general, the relative strength of the potential in each direction can be modified by changing the k-vector of the optical trapping beams. \par

We first consider the effect of the optical trapping on the axial modes of motion of the ion-crystal, that lie at the lower parts of the motional spectrum and are typically more prone to heating. When considering the joint motional mode structure of two identical ion-crystals, in two completely separate traps, we observe a degenerate mode structure, with each motional frequency appearing twice, and each motional mode being localized at one of the ion-crystals. Thus we intuitively expect that the motional spectrum of a well-decoupled optically segmented ion-crystal will form approximately degenerate bands. \par

\begin{figure}[!hbtp]
	\caption{Axial spectrum of an ion-crystal with N=231 ions, segmented into $S=6$ cells. Cells are separated by $B_A=3$ barrier ions and include $C+B_B=35$ ions each. The axial frequencies (blue) are shown for various trapping optical potentials, $\omega_\text{o.t.p}$ and are normalized by the characteristic Coulomb scale, $\nu$. For comparison, an additional axial spectrum of an independent ion-crystal with 35 ions is shown (dashed orange). For $\omega_\text{o.t.p}>\nu$ the axial spectrum of the optically segmented trap forms bands which are located at the frequencies of the independent cell. As $\omega_{\text{o.t.p}}$ increases, high-frequency bands form, which are due to the $B_A$ barrier ions. We highlight, $\omega_\text{o.t.p}/\nu\approx2.1$ (vertical line), used throughout the text. The inset is a zoom-in on the center-of-mass band (vertical axis in log-scale), showing that at $\omega_\text{o.t.p}=2\nu$ the segmented crystal's modes resemble the independent cell, implying that heating is dictated by the properties of a single cell rather than the whole, multi-cell, crystal.}
	\centering
	\includegraphics[width=1\linewidth]{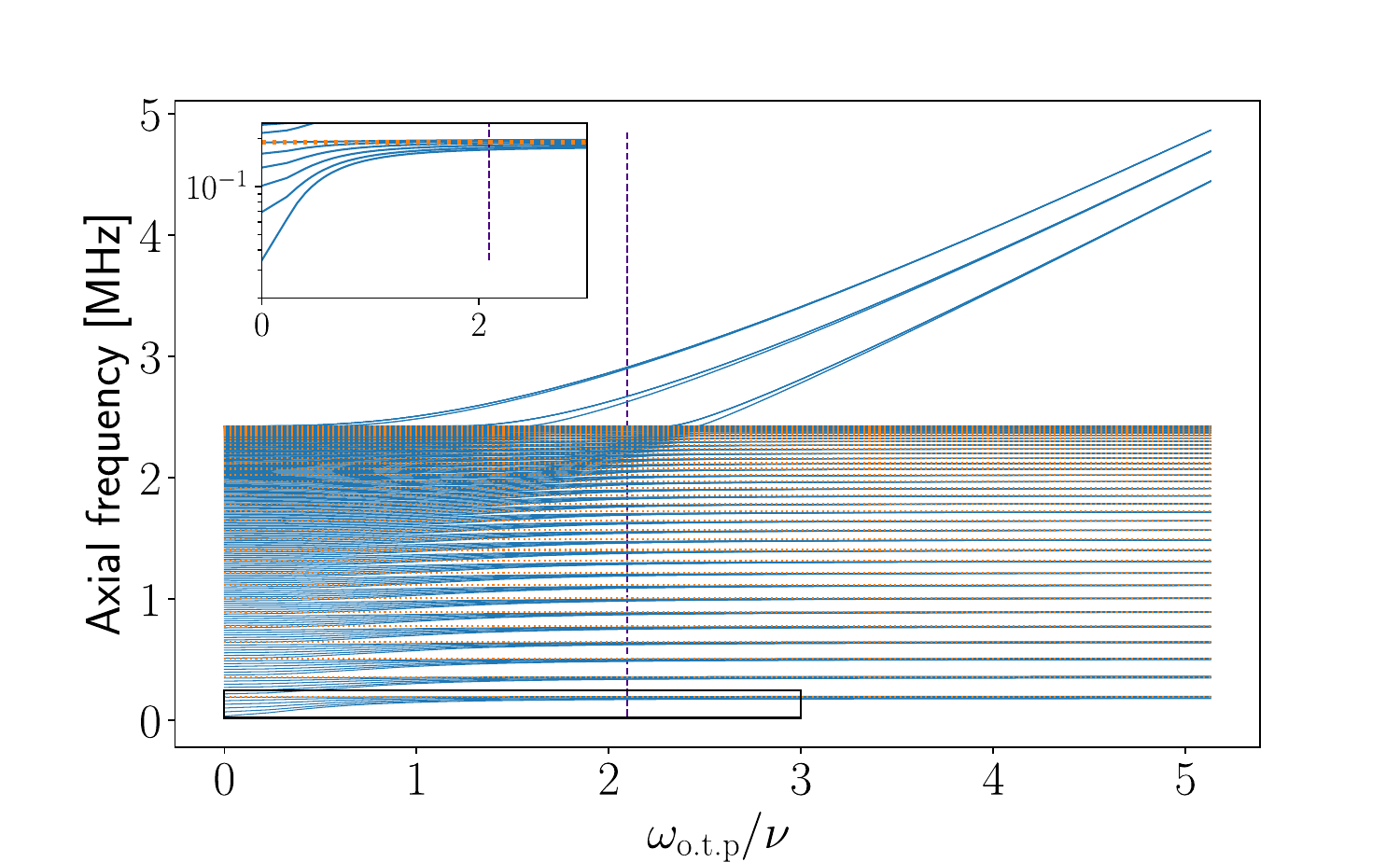}
	\label{figAxialSpectrum}
\end{figure}

Figure \ref{figAxialSpectrum} demonstrates this by showing the axial motional spectrum of a $N=231$ long ion-crystal for a varying optical trapping potential that segments the ion-crystal into $S=6$ cells, separated by $B_A=3$ barrier ions, such that each cell contains 35 ions (out of which $C=32$ are computational qubits and $B_B=3$ acts as barrier ions for segmentation configuration ‘B’). As the optical trapping potential increases the axial spectrum (blue) forms 35 bands of $S$ modes each. \par

Indeed the formation of bands heralds the decoupling of the motion of cells from each other, such that band frequencies resemble those of an independent, un-segmented, ion-crystal containing $C+B_B$ ions (orange), with each mode in the band being a superposition of local excitations of the corresponding mode of the independent cell. Thus, the width of each band marks the relevant rate in which a motional excitation traverses from one cell to another. That is, an excitation of a local mode in a single cell is composed of a superposition of all modes of the band, which then disperse at a time scale that is inversely proportional to the band width. \par

Figure \ref{figAxialSpectrum} further shows $B_A$ high-frequency bands, containing $S+1$ modes, associated with the motion of $\left(S+1\right)B_A$ barrier ions. Crucially these bands are separated from the ‘bulk’ ions such that their motion is essentially decoupled from the bulk. We exploit this fact below in order to implement mid-circuit measurements and cooling of the barrier ions. \par

\begin{figure}[!hbtp]
	\caption{Radial modes spectrum of an optically-segmented ion-crystal containing $N=231$ ions, and the resulting cell crosstalk. Due to the segmentation the mode frequencies (blue) form $C+B_B$ bands, which imitate the structure of an unsegmented independent cell (orange). Additional $B_A$  high-frequency bands associated with the barrier ions are formed as well. The bottom-right inset shows a zoom-in on the last four bulk bands. The bands are well separated and narrow in frequency. The top-left inset presents estimations of unwanted crosstalk between cells during simultaneous multi-qubit operations, mediated by the various bulk bands. These estimations are calculated by considering the system's spectrum, $\varepsilon_{\text{BW},b}$ (orange) or mode-structure considerations, $\varepsilon_{J,b}$ making use of the modes of the entire ion-crystal (blue lines), or only neighboring segments (green stars). Both predict a $\sim2.5\%$ crosstalk due to mainly long-wavelength motion of ions in the cells, associated with high bands. Mitigation of the crosstalk is presented in section IV in the main text.} 
	\centering
	\includegraphics[width=1\linewidth]{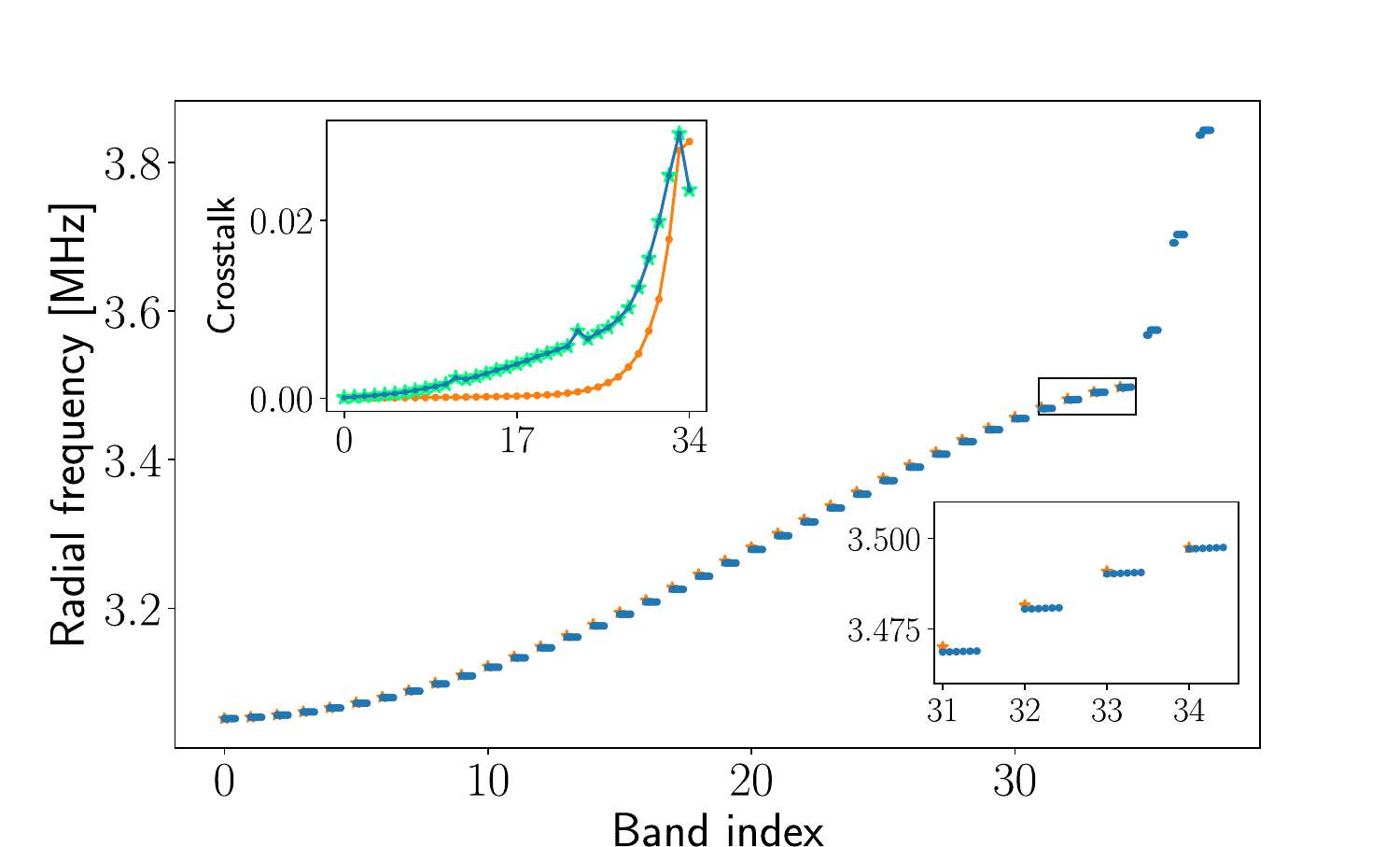}
	\label{figRadialSpectrum}
\end{figure}

Figure \ref{figRadialSpectrum} shows the transverse (radial) band structure of the same $N=231$ ions system with $\omega_\text{o.t.p}\approx2.1\nu$ (dashed vertical line in Fig. \ref{figAxialSpectrum}). Similarly to the axial direction, optical segmentation results in the formation of $C+B_B$ bulk-bands as well as additional $B_A$ bands associated with the motion of barrier ions (blue). The average frequencies of the bands resemble those of an independent cell (orange), also shown in the inset (bottom right) that presents a zoom-in on the last four computational qubit bands, showing narrow and well-separated bands. For transverse modes of motion the band index, $b$ (horizontal axis), ordered according to the band's frequency, is opposite to the mode's wave vector, i.e. large $b$'s represent long-wavelength modes of motion. \par

Transverse modes of motion are used in order to generate entanglement between computational qubits, by mediating qubit-qubit interaction via spin-dependent motion \cite{sorensen2000entanglement}. Typically the entanglement operation duration is inversely related to the frequency difference between adjacent modes of motion \cite{shapira2023fast}. \par

In our case, the frequency gap between bands, $\Delta\omega_b=\omega_b-\omega_{b-1}$, with $\omega_b$ the mean frequency of the modes of bulk bands, $b=1,...,C+B_B$, marks the typical interaction rate between ions within the same cell. By contrast, the coupling rate between different cells is determined by the bandwidth, $\text{BW}_b$, which is defined as the difference between the highest and lowest frequency modes in band $b$. Thus a simple estimation of unwanted crosstalk between cells during parallel entanglement operations is the ratio of these two rates - given by $\varepsilon_{\text{BW},b}=\frac{\text{BW}_b}{2\Delta\omega_b}$. The resulting estimate is presented in the inset (top left) of Fig. \ref{figRadialSpectrum} (orange), showing that high $b$ bands, associated with long-wavelength motion across cells, results in a higher crosstalk, at the few percent level. A more precise estimate of the crosstalk is also shown (blue line and green points) and discussed below. \par
 
Optically confined ions are not used as computational qubits since optical confinement significantly degrades their coherence due to spontaneous photon scattering and uncontrolled light-shifts. For example, we consider $^{40}\text{Ca}^+$ ions that are illuminated by a confining optical field at $400$ nm with a beam that has a diameter of $1\text{ }\mu\text{m}$, off-resonant with the $S\leftrightarrow P$ transition. In this setup, a harmonic confinement frequency of $\omega_\text{o.t.p}\approx2.1 \nu= (2\pi)*1.75$ MHz, used throughout our analysis, requires an optical power of $20$ mw illuminating each barrier ion. The resulting spontaneous photon scattering rate is approximately $3$ kHz. \par

This scattering rate, however, does not pose any fundamental challenges. Photons scattered from barrier ions are significantly detuned (by $10$'s of THz) with respect to the $S\leftrightarrow P$ transition in neighboring computational qubits, which thus acquire only a negligible decoherence rate due to this process. Moreover, because the barrier ions only participate weakly in bulk modes, the heating rate of these modes due to photon scattering is small, $\sim 10^{-2}$ quanta per second (see the SM \cite{SM}). It should be noted that the barrier ions themselves do heat as a result of this process; thus, barrier ions are required to be ground state cooled during each reconfiguration step, i.e. before they are released into the bulk. This point is further discussed in section VII. \par

We remark that it is possible to use different ion species for data qubits and barrier ions. This would allow to easily realize a high optical trapping frequency, by using a closer to resonance trapping field, that scatters many photons, without affecting data qubits. Cooling of such barrier ions is still required, yet becomes simpler as well.   

\section{Heating rates of optically-segmented ion-crystals}

A crucial criteria for successfully scaling-up ion-crystals is maintaining crystal stability. In the standard approach to trapped-ion crystals this challenge comes about as a heating rate of the crystal that scales unfavorably with $N$. Specifically, in un-segmented ion-crystals as $N$ increases, the frequencies of low-lying long-wavelength modes of motion decrease, and become more susceptible to electrical noise, leading to degradation of coherent operations, destabilization of the ion-crystal and its eventual melting. Utilizing optical trapping potentials generates cell-local heating rates, which only depend on the cell size, $C$, and not on the total number of ions $N$ (or $S$), potentially allowing for arbitrarily long ion-crystals.\par

This is shown graphically in Fig. \ref{figAxialSpectrum}, since for $\omega_\text{o.t.p}\geq2\nu$ the low-lying COM band is lifted and its frequency converges to the COM mode of an independent cell (inset), implying cell-dependent (and not $N$-dependent) heating rates. The COM modes remain gapped regardless of $N$. \par

We directly quantify the phonon excitation rate of the optically-segmented architecture. For a given motional mode, the phonon excitation due to the presence of some electric field noise, $\delta E(t,r)$,can be evaluated by Fermi's golden rule \cite{brownnutt2015ion}, yielding,\par
\begin{equation}
	\Gamma^{(k)} = \frac{e^2}{4m\hbar\omega_k}S_E^{(k)}(\omega_k)\label{eqGammaK},
\end{equation}
with $\omega_k$ the frequency of mode $k$, and $S_E^{(k)}(\omega_k)$ is the spectral density of the electric field noise. The mode's heating rate is $\hbar\omega_k\Gamma^{\left(k\right)}$. \par

The precise form of the spectral density function is a subject of much theoretical and experimental study \cite{lamoreaux1997thermalization,wineland1998experimental,king1998cooling,james1998theory,turchette2000heating,brownnutt2015ion,kalincev2021motional}, is system-dependent and its theoretical limits remains somewhat inconclusive. Yet it is generally agreed that it scales as, $S_E(\omega) \propto \omega^{-\alpha} D^{-\beta} T^\gamma$, with mode frequency $\omega$, ion-electrode distance $D$, and temperature, $T$, and $\alpha$, $\beta$ and $\gamma$ are scaling exponents. Motivated by many experimental results, here we assume that the noisy electric field is spatially uniform along the ion-crystal and that $\alpha=1$. \par

Using these consideration we obtain the system's excitation rate (see details in the SM \cite{SM}),
\begin{equation}
	\Gamma = \sum_k \Gamma^{(k)} \propto \sum_k  \frac{e^2}{4m\hbar\omega_k^{1+\alpha}} \sum_{i,j}A_{k}^{(i)} A_{k}^{(j)},\label{eqGamma}
\end{equation}
with $\Gamma^{(k)}$ the excitation rate of the $k$th mode and $A$ an orthogonal axial mode-matrix, such that $A_{k}^{(i)}$ is the participation of the $i$th ion in the $k$th axial mode of motion, obtained by analyzing the axial mode structure \cite{johanning2016isospaced,SM}. \par

Due to the summation over $k$, our formulation of $\Gamma$ in Eq. \eqref{eqGamma} is extensive, e.g. for two identical and independent copies of an ion-crystal, $\Gamma$ will be evaluated as twice the value of a single copy of the ion-crystal. Thus ideal decoupling between cells in the optically segmented ion-crystal is expected to show-up as proportionality of $\Gamma$ to $S$.\par

\begin{figure}[!hbtp]
	\caption{Axial excitation rate of optically segmented ion-crystals. The excitation rates (vertical axis), $\Gamma$, are shown for various values of cell sizes, $C$ (horizontal axis) and number of segments, $S$ (colors). We normalize $\Gamma$ by the excitation rate of an independent single cell, $\Gamma_\text{ind}$ and by the number of segments, $S$. All resulting normalized rates are of order unity, showing that the stability of the optically segmented ion-crystal is dictated by $C$, and is largely independent of $S$. We repeat this process but with only considering the lowest band of motional modes in the expression for $\Gamma$ (dashed), without varying $\Gamma_\text{ind}$. The resemblance between this result and the full expression of $\Gamma$ (solid) shows that the lowest-lying band is the dominant contributor to the heating. This result is intuitive as this band is a COM motional band, having a high-overlap with a uniform electric field.}
	\centering
	\includegraphics[width=1\linewidth]{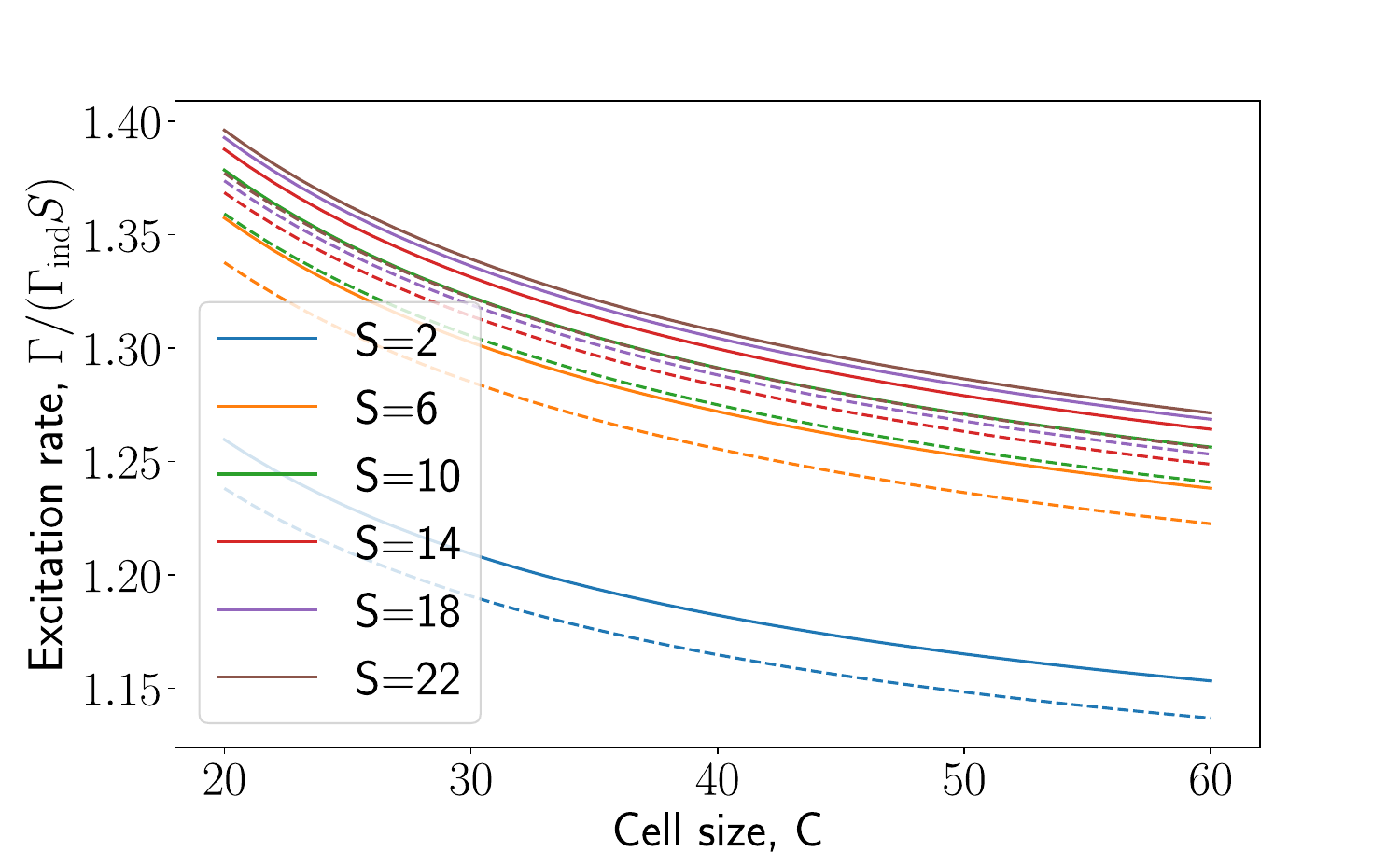}
	\label{figAxialHeating}
\end{figure}

Figure \ref{figAxialHeating} shows the predicted excitation rate of the optically segmented ion-crystal, due to Eq. \eqref{eqGamma}, for various values of the cell size, $C$ (horizontal axis), and number of segments, $S$ (colors). The barrier size is set to $B_A=3$  and each cell contains $C+B_B$ ions, with $B_B=3$. Excitation rates are calculated for ion-crystals reaching $N>1300$ ions. The figure shows, $\Gamma/\left(\Gamma_\text{ind}S\right)$ (vertical axis), such that the excitation rates are normalized by the number of segments, $S$, and by the excitation rate of an independent unsegmented ion-crystal containing $C+B_B$ ions, $\Gamma_\text{ind}$. We observe that the normalized excitation rates are all of order unity and converge towards 1 as $C$ increases, with a negligible dependence on the number of segments. Thus we conclude that the cell's stabilization is dominated by cell properties, and is specifically independent from other cells in the ion-crystal.\par

The main contribution to $\Gamma$ is given by the lowest frequency band, which is a COM motional band. This is intuitive as COM modes have the lowest motional frequencies, as well as a high overlap with the uniform electric field noise considered here, as compared to other modes which have negligible contributions. This is seen in the dashed lines of Fig. \ref{figAxialHeating} in which we calculate $\Gamma$ using exclusively this single band (while $\Gamma_\text{ind}$ is still calculated in full). \par

\section{Parallel multi-qubit logic and crosstalk mitigation}

In trapped-ion based quantum computers, entanglement gates are typically facilitated by spin-dependent forces acting on normal-modes of motion of the ion-crystal \cite{sorensen1999quantum,sorensen2000entanglement}. Thus adequate control of the normal-mode structure is crucial for scaling-up the trapped-ion quantum register size. Specifically, utilizing many modes of motion simultaneously enables programmable long-range interactions \cite{shapira2020theory,shapira2023programmable,shapira2023fast}, which are significant for efficient implementation of various quantum computational tasks \cite{breuckmann2021quantum,schwerdt2022comparing,bravyi2022constant,bassler2023synthesis,bravyi2023highthreshold,wu2023qubits,lu2023realization,shapira2023programmable}. However utilizing multiple modes of motion becomes increasingly challenging in large ion-crystals due to the overwhelming complexity of the required qubit drive, which comes about as a challenging optimization problem \cite{shapira2020theory}, as well as an apparent slow-down of the feasible gate duration \cite{shapira2023fast} which, in absence of segmentation, scales as $N^2$. Thus segmentation plays here a crucial role as well in generating independent large, yet manageable, computational qubit cells.\par

The heuristic estimation for crosstalk error, provided by $\varepsilon_\text{BW,b}$ above, can be improved by a detailed consideration of the transverse mode-structure. Specifically, based on the quadratic-form of qubit-qubit entanglement discussed thoroughly in \cite{shapira2023fast}, the coupling term between ions $c$ ($c^\prime$) in cells $s$ ($s^\prime$), mediated by band $b$, is scaled by the factor,
\begin{equation}
	J_b^{\left(c,s\right),(c^\prime,s^\prime)}=\sum_{m=1}^S R_{\left(b,m\right)}^{\left(c,s\right)}R_{\left(b,m\right)}^{(c^\prime,s^\prime)},\label{eqJeff}
\end{equation}
with $R$ an orthogonal radial mode-matrix (similar to $A$ above) such that $R_{\left(b,m\right)}^{\left(c,s\right)}$ is the participation of ion $c$ of cell $s$ in mode $m$ of band $b$ (specific examples of $J_b^{\left(c,s\right),(c^\prime,s^\prime)}$ are shown in the SM \cite{SM}). Using $J_b$ above, we form a more detailed crosstalk estimation, namely,
\begin{equation}
	\varepsilon_{J,b}=\frac{\max\limits _{c\neq c^{\prime},s\neq s^\prime}\left|J_{b}^{\left(c,s\right),(c^{\prime},s^{\prime})}\right|}{\max\limits _{c\neq c^{\prime},s}\left|J_{b}^{\left(c,s\right),(c^{\prime},s)}\right|},\label{eqCrosstalkJ}
\end{equation}
with $s,s^\prime$ maximized over all $S$ cells in the segmented ion-crystal and $c,c^\prime$ maximized over all $C$ computational qubits in each of the cells. In essence the estimation in Eq. \eqref{eqCrosstalkJ} compares the coupling between ions in the same cell ($s=s^\prime$) to coupling between ions in difference cells ($s\neq s^\prime$). This estimation is also shown in the inset (top-left) of Fig. \ref{figRadialSpectrum} (blue), exhibiting a similar behavior to that of $\varepsilon_{\text{BW},b}$, i.e. $\sim2.5\%$ crosstalk mainly due to long-wavelength motional excitations of the cells. We also evaluate separately the contribution of nearest-neighbor cells, $\varepsilon_{J,b}^\text{nn}$, which follows the same definition as $\varepsilon_{J,b}$ in Eq. \eqref{eqCrosstalkJ} but with $s^\prime=s\pm1$ in the numerator. This estimation is shown as well in the inset (top-left) of Fig. \ref{figRadialSpectrum} (green stars), showing an almost exact agreement with $\varepsilon_{J,b}$. Thus we conclude that the $\sim2.5\%$ crosstalk is mainly due to unwanted coupling between adjacent segmented cells. Further analysis shows that in general 2-3 barrier ions, each confined with an optical trapping frequency of $\omega_{\text{o.t.p}}\approx\omega_{\text{rad}}$ are sufficient for maintaining low crosstalk levels (see the SM \cite{SM}).\par

A few percent crosstalk level, reflected by both $\varepsilon_{\text{BW},b}$ and $\varepsilon_{J,b}$, marks an efficient decoupling between cells. Nevertheless, the remaining coupling is too high for high-fidelity quantum operations, and restricts performing simultaneous logical gates on distinct cells. However this crosstalk level is low enough such that it can be mitigated perturbatively in a scalable manner. Here we do so by relying on the large-scale fast (LSF) method, presented in Ref. \cite{shapira2023fast}. 

In essence, LSF makes use of a multi-tone drive in order to generate programmable multi-qubit gates, which naively involves solving a quadratically constrained optimization problem of dimension $N^2$, and generates unitary gates of the form $U=\exp\left(i\sum_{n,m=1}^N \varphi_{n,m}\sigma_x^{\left(n\right)}\sigma_x^{\left(m\right)}\right)$, with $\sigma_x^{\left(n\right)}$ a Pauli-$x$ operator acting on the $n$th qubit and $\varphi_{n,m}$ a target multi-qubit coupling matrix. This gate, together with arbitrary single qubit rotations realizes a universal gate set. \par

The operational approach of LSF is naturally adopted to our architecture, making the design of multi-qubit entanglement, simultaneously in all cells, fast and scalable. In particular, we use LSF for a ‘typical’ cell of the bulk of the ion crystal (see SM for details \cite{SM}) to obtain drive spectra corresponding to different target unitaries. These solutions could then, in principle, be applied to each cell within the ion crystal. Up to minor differences between the ‘typical’ and actual cells (to be corrected below), this would yield high-fidelity multi-qubit gates if applied on each cell separately. \par
However, these operations are meant to run in parallel and, as discussed above, the unwanted coupling between cells incurs a $\sim2.5\%$ crosstalk error - which is dominated by nearest-neighbor cells. Therefore we make use of an additional mitigation layer on top of LSF. Crucially, the nearest-neighbor structure of the crosstalk enables our mitigation technique to scale favorably, in terms of classical computation resources, and does not impose constraints on the ion-crystal size.\par 

Specifically, we perform an iterative optimization by linearizing the quadratic constraints associated with the target unitary around the current solution, with the initial solution given by LSF as discussed above. The resulting linear equations provide conditions that resolve crosstalk as well as inaccuracies of the LSF solution that originate from the assumption of a typical cell. We do so locally, i.e. the linear conditions are formed for only two adjacent cells at a time, generating $C\left(C-1\right)$ linear constraints that improve the LSF results and mitigate deviations of the cells from the ‘typical’ cell, and $3C^2$ linear constraints that mitigate crosstalk between cells. The iterative optimization is stopped when a target infidelity is reached. Crucially since the linear equations are local, relating only four adjacent cells at a time, they involve only $\mathcal{O}\left(C^2\right)$ linear constraints, allowing parallel optimization of the next four adjacent cells \cite{SM}.\par

An additional condition for a high-fidelity process is decoupling of the qubits from the modes of motion at the end of the entanglement operation. This is typically satisfied by forming constraints for each of the $N$ modes of motion independently. Here however such an approach will be a hurdle to scalability since it will be infeasible to satisfy a large $N$ number of constraints. Instead, we make use of only $C+B_{A,B}$ linear constraints of the typical cell, and supplement these with additional robustness properties that make the decoupling insensitive to the mode frequency inaccuracies, described in Ref. \cite{shapira2018robust}, such that each constraint enforces the decoupling of an entire band at once. The error in our gates is therefore expected to have a contribution from imperfect disentanglement of motion from qubits. \par

\begin{figure}[!hbtp]
	\caption{Performance of the crosstalk mitigation method as a function of optical trapping potential $\omega_\text{o.t.p}$, evaluated on the entanglement used to implement a simulation of the 3D rectangular Ising model in Fig. 
 \ref{figSimulation}. Top: Error per cell, using Eq. \eqref{eqInfid}, as a proxy for the infidelity of the entire multi-qubit operation. For $\omega_\text{o.t.p}\gtrsim2\nu$ the crosstalk compensated solution (dark-blue) shows a low, $<10^{-4}$, infidelity. As expected, the initial LSF solution, based on the effective typical cell (orange) shows a $\sim10^{-2}$ infidelity. The mitigation degrades in the weak optical confinement regime, due to long-range coupling between cells that is not considered in the crosstalk mitigation method. Indeed this degradation appears when considering long-range cells (dashed purple), and not seen when only considering nearest-neighbor and onsite coupling (dashed pink). Additional infidelity due to residual coupling to motional modes (dashed light-blue) exhibits a similar behavior. Bottom: The required total drive coupling, given in units of the lowest transverse mode frequency (bottom). The crosstalk compensated solution (dark-blue) only incurs a small, $\sim 10\%$, overhead on the initial LSF solution (orange).} 
	\centering
	\includegraphics[width=1\linewidth]{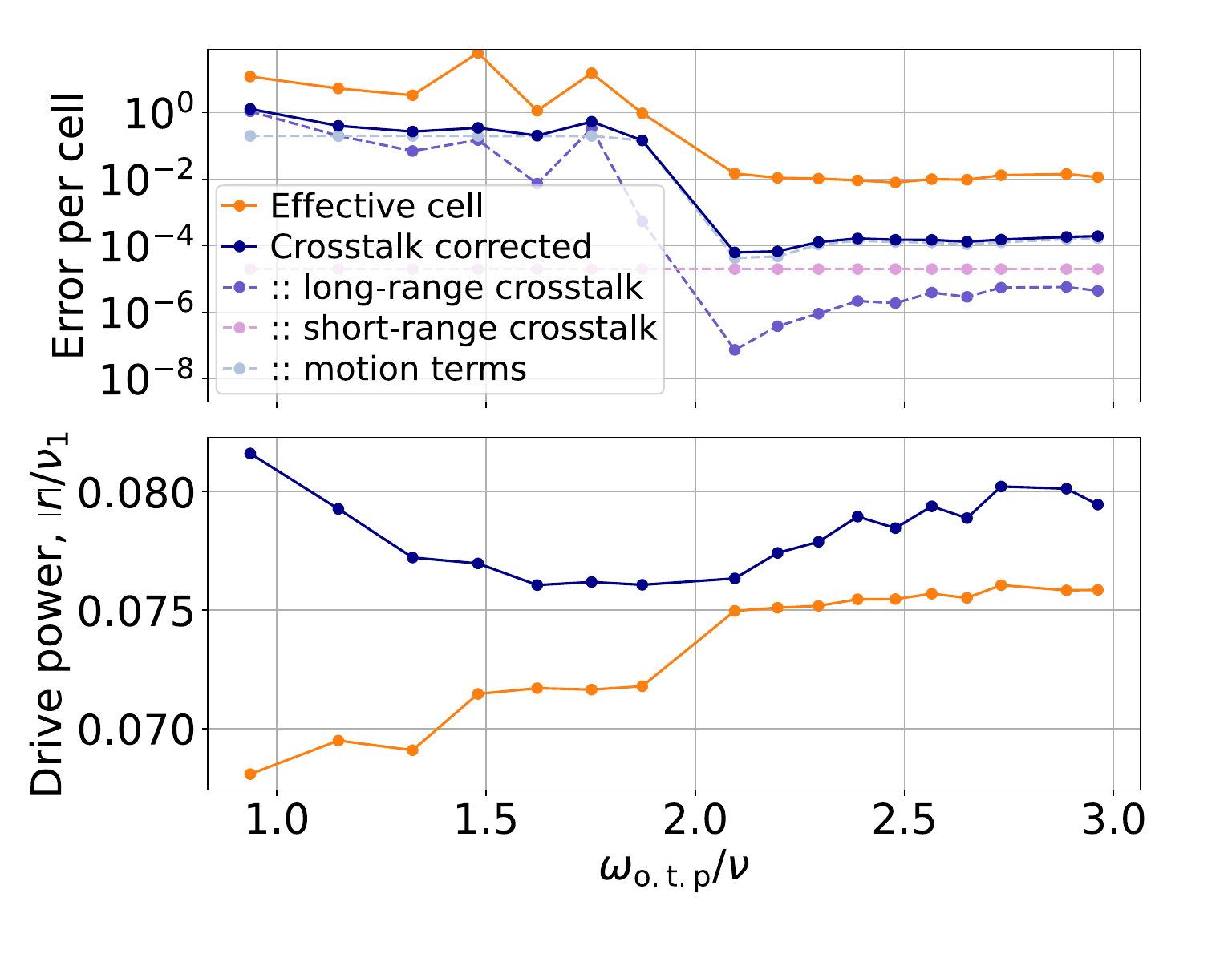}
	\label{figMitigationAnalysis}
\end{figure}

We study the efficacy of our crosstalk mitigation method by considering a segmented ion-crystal system used for a 3D Ising model simulation, as that shown in Fig. \ref{figSimulation} above, with $S=5$, $d=4$ and $B_A=B_B=3$. We design the control pulses that implement the required qubit-qubit couplings, for various optical trapping frequencies. The resulting operation infidelity, $I$, is evaluated, to leading order in the qubit-qubit coupling deviations and in residual ion displacements, as, 
\begin{equation}
	I=I_\text{short}+I_\text{long}+I_\text{motion} \label{eqInfid}
\end{equation}
such that $I_\text{short}$ approximates the infidelity due an erroneous qubit-qubit coupling within each cell and crosstalk to nearest-neighbor cells, $I_\text{long}$ is the infidelity due to qubit-qubit coupling between cells that are not nearest-neighbors and $I_\text{motion}$ approximates the infidelity due to residual coupling between the qubits and the phonon modes of motion. \par

Figure \ref{figMitigationAnalysis}(top) presents the infidelity of the initial LSF solution, based on the typical cell (orange) and the crosstalk corrected solution (dark-blue). It exhibits an apparent cross-over between two regimes at $\omega_{\text{o.t.p}}=2\nu$, at which point the transverse mode band structure is manifested. At low optical confinement the performance of uncorrected, typical-cell solutions (orange) is low, such that our leading-order estimation breaks-down. At strong optical confinement this solution generates an infidelity of $\sim10^{-2}$, as expected from our analysis above. However the crosstalk-corrected solution (dark-blue) successfully generates solutions that have an infidelity that is better (lower) than $10^{-4}$ for the entire multi-qubit operation, throughout the strong optical confinement regime.\par
	
We account for the different contributions to the infidelity of the crosstalk-corrected solution, given by Eq. \eqref{eqInfid}. We note that $I_\text{short}$ (dashed pink), has a negligible contribution, regardless of optical confinement, while $I_\text{long}$ (dashed purple) reaches a low values only at the strong confinement regime. This behavior is expected as our optimization actively corrects for on-site and nearest-neighbor terms, and ignores other terms. We note that $I_\text{long}$ reaches a minimum at $\omega_\text{o.t.p}=2\nu$, and is slightly increased at higher optical trapping frequencies. This behavior is accounted by an effective screening of barrier ions that occurs when barrier bands are close to resonance with the bulk bands (see the SM \cite{SM}). Nevertheless, our analysis implies that strong optical confinement successfully decouples next nearest-neighbor (and further) cells, removing the need to take the entire ion-crystal into account, and enabling scalability. Similarly, due to a well-formed band structure, the motional infidelity, $I_\text{motion}$ (dashed light-blue), becomes small at strong optical confinement, and in general dominates the overall infidelity.\par

Figure \ref{figMitigationAnalysis}(bottom) shows the total drive coupling, i.e. Rabi frequency, required to drive the initial LSF solution (orange) and the crosstalk corrected solution (dark-blue), given in terms of the frequency of the lowest radial mode, $\nu_1$, as a characteristic scale. The compensated solution incurs a small overhead of $\sim10\%$ on the initial solution.\par

We remark that elements of our architecture and crosstalk compensation method can also be implemented by global driving of entire cells using similar techniques to those in Refs. \cite{shapira2020theory,shapira2023fast}. Moreover by employing the established methods in Refs. \cite{haddadfarshi2016high,webb2018resilient,leung2018robust,milne2020phase,blumel2021efficient,shapira2023robust,kang2023designing}, we can impose additional robustness constraints on each gate drive in order to reduce the sensitivity of the gate and the crosstalk compensation, to fluctuations in trap parameters as well as other sources of noise.  

\section{Mid-circuit measurements}

The ability to perform mid-circuit projective measurements, and apply coherent feedback based on the measured results, is at the heart of QEC, as well as additional central quantum computational tools \cite{griffiths1996semiclassical,rattew2021efficient,lu2022measurement}. In trapped-ion based systems, projective state detection is typically performed by state-dependent florescence. This poses a technical challenge, as the scattered photons usually heat-up the ion-crystal due to photon recoil. Furthermore the photon may be resonant with neighboring ions and scattered by them, resulting in decoherence of the quantum state of the entire system. Thus, so far, mid-circuit measurements have been implemented in trapped ions either in small ion-crystals, which are well separated from other parts of the quantum register, by ‘shelving’ all computational qubits to non-resonant atomic states, or by using two different atomic species spectraly separating logical and incoherent operations \cite{schindler2011experimental,negnevitsky2018repeated,ryan2021realization,manovitz2022trapped}.\par

Here however, optically confined ions are well-suited for mid-circuit measurements, as their motion is separated to independent bands (top part of the spectra in Figs. \ref{figAxialSpectrum} and \ref{figRadialSpectrum}), thus photon scattering heating remains local and does not heat the other crystal modes. Furthermore, the optical trapping also substantially light-shifts the atomic transition lines of the ions, specifically the photon emission lines, such that secondary photon scattering from neighboring computational qubit ions is largely suppressed. Thus, no physical shuttling of the ions into dedicated measurement regions is required, nor the use of multiple ion-species. \par

Nevertheless, measured ions may heat substantially, thus each measurement step is followed by mid-circuit cooling. Only the modes of optically confined ions must be cooled this way, leaving the information encoded in computational qubit ions unaffected. Mid-circuit cooling can be done on all optically confined ions in parallel, and is expected to occupy a minor part of the overall circuit duration (see the SM for further detail \cite{SM}). \par

\begin{figure}[!hbtp]
	\caption{Top: Schematic of mid-circuit measurement protocol, showing three types of involved ions: computational qubit ions (top, white circle), barrier ions ($B_\text{B}$) to be measured (middle, white circle), and barrier ions ($B_\text{A}$) that are currently optically confined (bottom, purple filled circle). Operations are ordered from left to right. Data is encoded on the computational qubit (e.g. $U_{s,n}^{\left(A\right)}$), and one of the qubit states of the ion is ‘shelved’ to a non-fluorescing state using local control. Next an intermediate optical segmentation configuration is used, confining both A and B barrier ions, and separating their motion from the computational qubits. The state of the barrier ions is measured and then reset (‘Prepare’). Lastly the optical segmentation configuration is switched. Bottom: Relevant atomic levels of $^{40}\text{Ca}^+$ ions, showing $4S_{1/2}$ the ground state manifold, the metastable $4D_{5/2}$ manifold and the short-lived $4P_{1/2}$ manifold. The $S\leftrightarrow P$ transition varies between the no-optical confinement (left) and optically confined (right) cases, allowing to separate the two cases spectrally (see main text). The main optical field used in the protocol are shown, namely at $729$ nm (red), coupling the $S\leftrightarrow D$ levels, at $400$ nm (light-blue), generating Raman transition between qubit states as well as additional local light-shifts, and at $397$ nm (purple) coupling the $S\leftrightarrow P$ levels.} 
	\centering
	\includegraphics[width=0.8\linewidth]{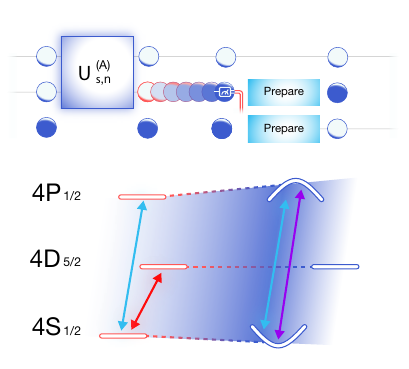}
	\label{figMidcircuit}
\end{figure}

We present mid-circuit measurements in conjunction with the ion-crystal's segmentation reconfiguration, as the two processes naturally combine, though each of these can also be performed independently. The protocol consists of measuring the qubit states of some target ions, without incurring decoherence on computational qubits, or heating of bulk modes, after which we cool, re-initialize the measured ions and reconfigure the ion-crystal's segmentation.  \par

Figure \ref{figMidcircuit} (top) illustrates these general steps involved in performing mid-circuit measurements. For simplicity we consider the required operations on three types of ions in the ion-crystal, i.e. computational qubit ions (top, white circles), barrier ions to be measured (middle, white circle), and optically confined barrier ions (bottom, purple filled circles). The measurement method can be performed in parallel throughout the entire ion-crystal on all required ions according to these roles. In our architecture ions may be optically confined specifically for the purpose of mid-circuit measurements, without necessarily having an essential role in segmentation of the ion-crystal, as showcased in the application of a QEC surface code, above. \par 

For concreteness we consider a specific realization, in trapped $^{40}\text{Ca}^+$ ions. Furthermore we consider local control over a wavelength $\sim400$ nm, which couples the two qubit states defined in the $4S_{1/2}$ ground state manifold via Raman transitions mediated by the short lived $4P$ manifolds. This local control, acting independently on all ions, is primarily used for generating single qubit rotations and multi-qubit programmable gates; however here it is utilized as a means to ‘localize’ global control fields using light-shifts, detailed below.  \par

The main atomic levels used for standard qubit operation are presented in Fig. \ref{figMidcircuit} (bottom) in the case of no optical confinement (left, red lines) and with optical confinement (right, blue lines). The ground state manifold is coupled to the narrow meta-stable $4D_{5/2}$ manifold via a $729$ nm field (red arrow). In the case of no optical confinement, this transition can be addressed on specific individual ions by light-shifting the $4S_{1/2}$ manifold of a target ion by using the locally controlled $400$ nm field (blue arrow). For these purposes a small, e.g. few MHz, light-shift suffices for individual ions addressing, while maintaining a low decoherence rate. On the other hand, optical confinement strongly light-shifts the $4S_{1/2}\leftrightarrow 4P_{1/2}$ level (right), making the $729$ nm coupling off resonant. In principle, this strong shift can be taken into account, however since the $S\leftrightarrow D$ transition is narrow, it imposes stringent requirements on the intensity stabilization of the optical confinement. Here we simply do not make use of the $S\leftrightarrow D$ transition for optically confined ions, relaxing the requirements for intensity stabilization. Raman transitions within the qubits subspace, mediated by the $400$ nm field, are negligibly affected by optical confinement, and exclusively in terms of their Rabi frequency. \par

An additional locally-applied field resonant with the light-shifted $S\leftrightarrow P$ transition at $397$ nm is used (purple arrow). Since the $P$ level is broad, the required stability of the optically confining fields in order to maintain resonance is attainable. The combination of local $400$ nm and $397$ nm fields, together with standard repump fields (not shown) accommodates for all the necessary steps - i.e., state detection, various cooling techniques, and state preparation. We note that the technical requirements for the $397$ nm field are less demanding than those of the optical confinement and gate drive fields at $400$ nm; the necessary power per ion is far lower, independent spectral control is not required, and the focal spot can generally be larger (by illuminating adjacent ions within the same beam).\par

The mid-circuit measurement protocol shown in Fig. \ref{figMidcircuit} is carried out by first encoding some information on ion to be measured (middle) via the entanglement operations detailed above ($U_{s,n}^{\left(A\right)}$), e.g. parity information of a plaquette shown in Fig. \ref{figQEC}. Then, one of the qubit states of the ion to be measured is ‘shelved’ to the D manifold via a combination of $400$ nm light-shift and $729$ nm coupling, making this operation local on the ion. \par

Next, the segmentation configuration is changed to an intermediate setting where all barrier ions are illuminated. This separates the motion of measured ions from the computational qubits' motion and light-shifts their $S\leftrightarrow P$ transition. At this step, some ions may fully populate the $D$ level, in which case they are not optically confined; thus existing confinement of other ions (bottom) cannot be relaxed. \par
State detection is then performed using the light-shifted $397$ nm field, followed by qubit reset and preparation using a combination of the $397$ nm and $400$ nm fields. During the detection and preparation steps, the optically confined ions scatter $397$ nm photons. The scattered photons are substantially detuned, by approximately $0.8$ GHz, from the absorption lines of the computational qubit ions, having a linewidth of few tens of MHz, thus negligibly influencing their state. Nevertheless the state of neighboring computational qubits, located several micrometers apart, may be further protected by shelving both of their qubit states to the $D$ level (at the shelving step). \par

The measured ions are required to be ground state cooled in the ‘prepare’ step, which can be done on all optically confined ions in parallel. Incidentally, this cooling step also dissipates the heat that accumulates on barrier ions during the circuit due to off-resonant photon scattering and due to measurements. \par

During exposure to $397$ nm light, the measured ions inevitably spend some portion of time in the excited $P$ state, in which the optical potential is anti-trapping. Moreover, transitions to the $P$ state may introduce heating due to the fluctuating dipole force \cite{Hutzler_2017,Alt_thesis}. The heat induced by these processes yields an excitation of $\sim5$ motional quanta of their localized mode, keeping them in a manageable regime, from which they can be recooled back to their motional ground state, detailed in the SM \cite{SM}. \par

Lastly, the segmentation configuration can be changed, switching the roles of computational and barrier ions. Although the barrier ions are cooled before returning to the bulk, care must be taken to reconfigure the optical potentials slowly enough so as not to excite any motion of the bulk modes. As detailed in the SM \cite{SM}, this condition is satisfied as long as the reconfiguration time is large compared to an oscillation period of the ion trap; in practice, a reconfiguration time of $\sim 30\mu\text{s}$ easily meets this requirement. Overall the duration of the mid-circuit measurement, preparation, and reconfiguration steps combined is generally small compared to the multi-qubit gate time.

\section{Summary}

Here we propose a scalable architecture for quantum computing; it is based on a large register of trapped-ion qubits together with dynamically operated optical potentials. Our proposed architecture circumvents the two most prominent challenges in working with ever-larger ion-crystals – prohibitively high heating rates and spectral crowding of the ions’ motional modes. It does so by effectively segmenting an arbitrarily large trapped-ion crystal into several independent segments of a manageable size. Connectivity across the full trapped-ion crystal is enabled by rapidly reconfiguring the optical potentials. The optical potentials further enable mid-circuit measurements of the confined ions, followed by classical feedback.\par
 
The utility of this architecture is emphasized when combined with a method for programmable multi-qubit entangling gates, such as that proposed in Ref. \cite{shapira2023fast}. We have used this method to numerically study the application of independent multi-qubit unitaries on each cell in parallel and in a scalable manner. Moreover we have extended this method to enable arbitrarily good compensation of crosstalk errors that arise between adjacent segments. \par

Our architecture requires modest hardware resources and makes use of well-established experimental techniques; it is thus an ideal platform for quantum simulation, near-term quantum computation, as well as ultimately for fault tolerant quantum computing.\par

\begin{acknowledgments}
	This work was supported by the Israel Science Foundation Quantum Science and Technology (Grants 2074/19, 1376/19 and 3457/21).
\end{acknowledgments}

\bibliographystyle{unsrt}
\bibliography{reference}

\onecolumngrid

\section{Supplemental material}

\subsection{Mode structure and coupling calculations}

Mode structure, i.e. motional mode frequencies and mode participation matrices are generated by diagonalizing a classical Hamiltonian. Specifically we consider an isospaced RF trapped-ion crystal, of ions with mass $m$ and charge $e$. Optical trapping potentials are taken into account by incorporating an additional on-site potential term for illuminated ions. In all of our examples in the main text and below we have considered equally-spaced trapped $^{40}\text{Ca}^+$ ions, with an inter-ion distance of $5\text{ }\mu\text{m}$ and a $400\text{ nm}$ optical field that generates the additional confinement and segmentation.

We assume the ion-crystal forms a stable linear chain, such that axial and transverse modes can be considered independently. Thus, following Ref. \cite{johanning2016isospaced}, for axial modes we write the secular matrix,
\begin{equation}
	V_{n,m}^{\text{axial}}=\begin{cases}
		-2\frac{eE_{\text{n.n}}}{d\left|n-m\right|^{3}} & n\neq m\\
		4\zeta\left(3\right)\frac{eE_{\text{n.n}}}{d}+m\omega_\text{o.t.p}^{2}b_{n} & n=m
	\end{cases},\label{eqSupSecAx}
\end{equation}
with $b_n=1$ ($b_n=0$) if ion $n$ is (not) illuminated by an optical potential, $d$ the distance between ions, $\zeta\left(3\right)\approx1.202$ is Apéry's constant, $E_\text{n.n}=\frac{e}{4\pi\epsilon_0 d^2}$ is the electric field strength created by nearest-neighbor ions and $\epsilon_0$ is the vacuum permittivity. The eigenvalues of $V^\text{axial}$ are $\frac{1}{2}m\omega_\text{a}^2$, with $\omega_\text{a}$ the axial motional mode frequencies. The eigenvectors are normal-modes of motion associated with the corresponding frequency, designated in the main text as the mode-matrix $A$. 

Similarly, transverse (radial) modes are calculated by diagonalizing the matrix,
\begin{equation}
	V_{n,m}^{\text{radial}}=\begin{cases}
		\frac{eE_{\text{n.n}}}{d\left|n-m\right|^{3}} & n\neq m\\
		-2\zeta\left(3\right)\frac{eE_{\text{n.n}}}{d}+m\left(\omega_{\text{rad}}^{2}+\omega_{\text{o.t.p}}^{2}b_{n}\right) & n=m
	\end{cases},\label{eqSupSecRad}
\end{equation}
with $\frac{1}{2}m\omega_\text{rad}^2$ the radial trapping potential generated by the RF trap, which in the analysis presented here is fixed at  $\omega_\text{rad}=(2\pi)*3.5$ MHz. Figures \ref{figAxialSpectrum} and \ref{figRadialSpectrum} of the main text  are generated by computing and diagonalizing the matrices in Eqs. \eqref{eqSupSecAx} and \eqref{eqSupSecRad}.

The resulting transverse mode-matrix $R$, is the orthogonal matrix that diagonalizes the secular matrix in Eq. \eqref{eqSupSecRad} and is used in the main text to estimate crosstalk between segments. This is realized by considering the mode-dependent qubit-qubit coupling (Eq. \eqref{eqJeff} of the main text),
\begin{equation}
	J_b^{\left(c,s\right),(c^\prime,s^\prime)}=\sum_{m=1}^S R_{\left(b,m\right)}^{\left(c,s\right)}R_{\left(b,m\right)}^{(c^\prime,s^\prime)},\label{eqSupJeff}
\end{equation}

\begin{figure}[!hbtp]
	\caption{Band-dependent qubit-qubit coupling, $J_b$. Horizontal and vertical axes are indices of qubits within the entire ion-crystal (excluding $B_A$ barrier ions). Couplings are shown in a symlog scale (colors represent a logarithmic scale in both the positive and negative directions from the origin, with an interval between $\pm10^{-6}$ that is linearly scaled). The optical segmentation generates a block-diagonal structure, that signifies strong coupling within each cell and negligible coupling between qubits in different cells. Couplings due to three bands are shown, the ‘zig-zag’ $b=0$ band (left), the COM $b=34$ band (right) and an intermediate band (center). Clearly crosstalk, seen as coupling outside of the diagonal blocks, is more relevant for high $b$, long wavelength modes.}
	\centering
	\includegraphics[width=0.75\linewidth]{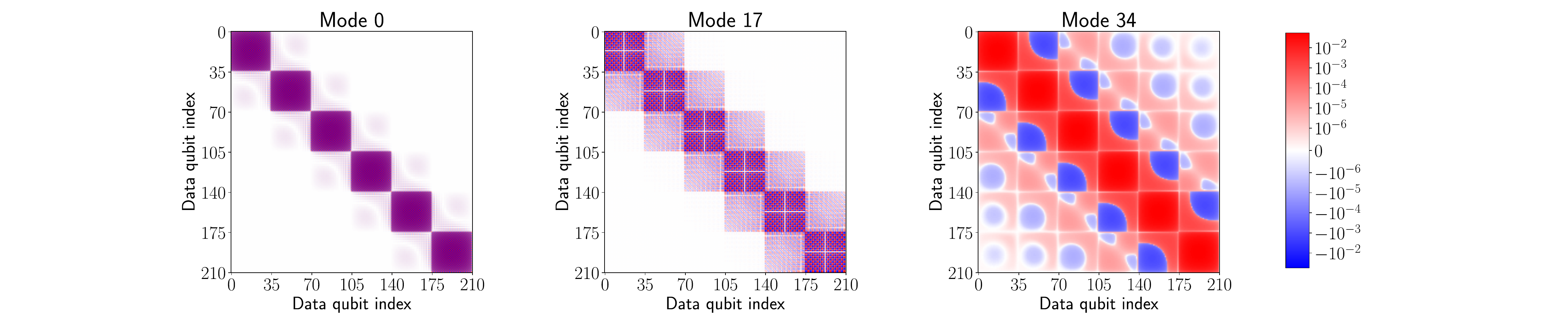}
	\label{figSupRadialModes}
\end{figure}

Figure \ref{figSupRadialModes} highlights the resulting structure of the $J_b$'s for several bands, with the horizontal and vertical axes representing ion indices within the ion-crystal (excluding $B_A$ barrier ions). Specifically the ‘zig-zag’, $b=0$, band (left) the COM, $b=34$, band (right) and an intermediate, $b=17$,  band (middle). Couplings are shown in a symlog scale (colors represent a logarithmic scale in both the positive and negative directions from the origin, with an interval between $\pm10^{-6}$ that is linearly scaled). All of the presented coupling maps show considerable inter-cell coupling and negligible crosstalk, i.e. coupling between ions in different segments, with nearest-neighbor segments constituting the most relevant correction. Furthermore long-wavelength, i.e. high $b$, modes generate more crosstalk, as is also shown in Fig. \ref{figRadialSpectrum} of the main text.

\subsection{Motional excitation rate calculations}

In order to evaluate the excitation rate of a given axial mode of motion we recall Eq. \eqref{eqGammaK} of the main text, i.e. in presence of some electric field noise, $\delta E(t,r)$, the excitation rate for a given motional mode is \cite{brownnutt2015ion},\par
\begin{equation}
	\Gamma^{(k)} = \frac{e^2}{4m\hbar\omega_k}S_E^{(k)}(\omega_k)\label{eqSupGammaK},
\end{equation}
with $\omega_k$ the frequency of mode $k$ and $S_E^{\left(k\right)}\left(\omega_k\right)$ the spectral density of the electric field noise, given by,
\begin{equation}
	S_E^{\left(k\right)}\left(\omega_k\right)=2\sum\limits_{i,j}\int d\tau\langle\delta E_k^i\left(\tau\right)\delta E_k^j\left(0\right)\rangle e^{-i\omega_k\tau}.\label{eqSupSe}
\end{equation}
Here $\delta E_k^i\left(\tau\right)$ is the projection of $\delta E$ on the $k$th mode of motion at the position of ion $i$, i.e. $\delta E_k^i\left(t\right)=\delta E(t,r_i)A_k^{\left(i\right)}$ and $r_i$ is the position of the $i$th ion. 

We assume that the leading order contribution to $\delta E$ is spatially uniform and neglect all other contributions. Furthermore, motivated by theoretical and experimental results we assume that $S_E^{\left(k\right)}\propto\omega_k^{-\alpha}$, with $\alpha$ here set to 1. 

This considerably simplifies $\Gamma^{\left(k\right)}$ and enables us to obtain the total axial excitation rate shown in Eq. \eqref{eqGamma} of the main text, up to a proportionality factor,
\begin{equation}
	\Gamma\propto\sum\limits_{k}\Gamma^{\left(k\right)}\propto\sum\limits_{k}\frac{e^2}{4m\hbar\omega_k^2}\sum\limits_{i,j}A_k^{\left(i\right)}A_k^{\left(j\right)}.\label{eqSupGamma}
\end{equation}

Crucially, the effect of optical segmentation comes about spectrally in the formation of bands, such that the summation on $k$ in Eq. \eqref{eqSupGamma} can be decomposed as,
\begin{equation}
	\Gamma\approx\sum\limits_{b=1}^{B+C}\frac{e^2}{4m\hbar\omega_b}\sum\limits_{s=1}^S\sum\limits_{i,j}A_{b,s}^{\left(i\right)}A_{b,s}^{\left(j\right)}+\mathcal{O}\left(\frac{\text{BW}_b}{\Delta\omega_b}\right),
\end{equation}
with $A_{b,s}^{\left(i\right)}$ a re-arrangement of $A$ according to the band structure and the error term due to the bandwidth of the bands, $\text{BW}_b$ and the gap between bands, $\Delta\omega_b$, which is assumed to be small. From a mode-structure point of view, the optical segmentation comes about as an independent motion between adjacent cells (shown in Fig. \ref{figSupRadialModes}), thus the sum on $s$ restricts $i$ and $j$ to the same cell, leading to a linear scaling with $S$.

\subsection{Crosstalk mitigation method}
We provide further details on our approach to mitigating crosstalk between cells in the optically-segmented ion-crystal. In order to do so we recall some basic facts about the operation of the LSF method, which is discussed in detail in Ref. \cite{shapira2023fast}. We assume the ions are driven by local fields that, in the spectral domain, all have the same $M$ tone pairs, but can vary in the amplitude of each pair. Thus the drive is described by a set of vectors $\textbf{r}_n$ such that $\left(\textbf{r}_n\right)_m$ is the amplitude of the $m$th tone pair that drives the $n$th ion. This choice of degrees of freedom allows to generate a bi-partite multi-qubit entanglement operation of the form,
\begin{equation}
	U=\exp\left(\sum\limits_{n,m=1}^N\varphi_{n,m}\sigma_x^{\left(n\right)}\sigma_x^{\left(m\right)}\right),\label{eqSupU}
\end{equation}
with $\varphi_{n,m}$ a completely controlled ‘target’ operation and $\sigma_x^{\left(n\right)}$ a Pauli-$x$ operator acting on the $n$th ion. 

Indeed, generating $U$ can be mapped to finding solutions (of the $\textbf{r}_n$'s) to the quadratic constraints,
\begin{equation}
	\textbf{r}_n^T A_{n,m} \textbf{r}_m=\varphi_{n,m}\quad 1\leq n<m\leq N,\label{eqSupQuad}
\end{equation}
with $A_{n,m}$ a set of real-valued $M\times M$ symmetric matrices that quantify the coupling between ions $n$ and $m$ due to the drive's spectral components. Specifically these coupling matrices are given by,
\begin{equation}
	A_{n,m}=\sum\limits_{j=1}^N R_j^{\left(n\right)}R_j^{\left(m\right)} A_j,\label{eqSupAnm}
\end{equation}
with $A_j$ a set of real-valued $M\times M$ symmetric matrices that quantify the coupling between the drive's spectral components and mode $j$ \cite{shapira2020theory}. Typically we will be interested in solutions to small norm $\textbf{r}_n$'s that satisfy Eq. \eqref{eqSupQuad}, which constitutes a NP-hard optimization problem.

In LSF we make use of a given non-trivial solution, $\textbf{r}_n=\textbf{z}$ for all $n=1,..,N$, that generates  the target $\varphi_{n,m}^{\left(0\right)}=0$ for all $n$ and $m$, coined the zero-phase solution (ZPS). Then, solutions to other ‘full’ arbitrary targets can be efficiently converted from $\boldsymbol{z}$. Crucially the ZPS does not depend on the ion index, $n$, and thus can be found directly by considering the $A_j$'s (instead of the $A_{n,m}$'s), reducing the number of quadratic constraints that are required to be satisfied.

We utilize these concepts and apply them to the optically-segmented ion-crystal. Generally we consider a ‘typical’ effective cell as an independent ion-crystal and find a ZPS for it. We then convert the ZPS to solutions of the full targets, which reflect the required operations on the different cells, i.e. $U_{s,n}^{\left(A,B\right)}$ in Fig. \ref{figMain} of the main text. By simply setting these solutions as the drive of the optically segmented ion-crystal we obtain entanglement gates of the form of Eq. \eqref{eqSupU}, up to $\sim10^{-2}$ infidelity, which arises mainly due to crosstalk between adjacent cells, but also due to using a typical cell system (e.g. cells at the edge of the ion-crystal are slightly different than in the bulk). 

To construct a typical cell system of the ion-crystal, we consider a fictitious system that has $C+B_B$ equally spaced ions with corresponding transverse motional modes. The mode frequencies are constructed from the average frequency of each band in the motional spectrum band structure of the segmented ion-crystal. The mode participation matrix is constructed by similarly averaging the participation of each ion in each mode, partitioned to cells. The segmented ion-crystal has $S$ modes per band (and not a single mode), which needs to be taken in to account by the typical cell system. Indeed we do so by writing our drive degrees of freedom in a way that is resilient to small inaccuracies in the motional frequency \cite{haddadfarshi2016high,shapira2018robust,webb2018resilient}. 

This construction, along with other fixed system parameters such as the entanglement gate time, ion type etc., is sufficient in order to construct $A_j$'s and $A_{n,m}$'s of a typical cell, and to generate ZPS of it. We remark that we do not make use of tones that lie within bands since these will generate a differential coupling between the different modes within a band, and are not well approximated by the typical cell system.

We use LSF to convert the typical cell ZPS to different target cell solutions for the various cells, and obtain a set of amplitudes, $\boldsymbol{r}_{s,c}^{\left(0\right)}$ corresponding to the drive of qubit $c$ in cell $s$. The qubit-qubit coupling between ions $c$ ($c^\prime$) in cell $s$ ($s^\prime$), ‘inherited’ from the typical LSF solution, can be evaluated as, $\varphi_{\left(s,c\right),\left(s^\prime,c^\prime\right)}^{\left(i\right)}=\left(\textbf{r}_{s,c}^{\left(i\right)}\right)^T A_{\left(s,c\right),\left(s^\prime,c^\prime\right)}\boldsymbol{r}_{s^\prime,c^\prime}^{\left(i\right)}$, with designating the LSF solution as $i=0$. 

The resulting couplings are compared to the intended target, $\Delta\varphi_{\left(s,c\right),\left(s^\prime,c^\prime\right)}^{\left(i\right)}=\varphi_{\left(s,c\right),\left(s^\prime,c^\prime\right)}^{\left(i\right)}-\varphi_{\left(s,c\right),\left(s^\prime,c^\prime\right)}^t$, with the ‘t’ inscription referring to the required ideal target dictated by the intended unitary operator, such that, $\varphi_{\left(s,c\right),\left(s^\prime\neq s,c^\prime\right)}^t=0$. Nearest-neighbor cell crosstalk errors are given by, $\Delta\varphi_{\left(s,c\right),\left(s\pm1,c^\prime\right)}^{\left(i\right)}$, and target inaccuracies are given by, $\Delta\varphi_{\left(s,c\right),\left(s,c^\prime\right)}^{\left(i\right)}$.

Since the total infidelity is small, crosstalk and target inaccuracies can be mitigated by linearizing the quadratic constraints in Eq. \eqref{eqSupQuad} and iteratively improving the resulting fidelity. Specifically we consider an iteration of the form $\boldsymbol{r}_{s,c}^{\left(i+1\right)}=\boldsymbol{r}_{s,c}^{\left(i\right)}+\boldsymbol{\delta}_{s,c}^{\left(i+1\right)}$, starting from $i=0$. We construct a set of linear equations for the correction $\boldsymbol{\delta}^{\left(i+1\right)}$. In order to make this technique scalable we want to avoid considering all $N$ ions in the ion-crystal in the same set of linear equations. This is made possible by the fact that crosstalk is dominated by coupling to nearest-neighbor cells (see Fig. \ref{figRadialSpectrum} in the main text and Fig. \ref{figSupRadialModes}).

Specifically we focus on two adjacent ‘target’ cells, $s$ and $s+1$, and derive linear equations for them. These are,
\begin{align}
	\left(\boldsymbol{\delta}_{s,c}^{\left(i+1\right)}\right)^T A_{\left(s,c\right),\left(s,c^\prime\right)}\boldsymbol{r}_{s,c^\prime}^{\left(i\right)}+\left(\boldsymbol{\delta}_{s,c^\prime}^{\left(i+1\right)}\right)^T A_{\left(s,c\right),\left(s,c^\prime\right)}\boldsymbol{r}_{s,c}^{\left(i\right)}&=\Delta\varphi_{\left(s,c\right),\left(s,c^\prime\right)}^{\left(i\right)}\\
	\left(\boldsymbol{\delta}_{s+1,c}^{\left(i+1\right)}\right)^T A_{\left(s+1,c\right),\left(s+1,c^\prime\right)}\boldsymbol{r}_{s+1,c^\prime}^{\left(i\right)}+\left(\boldsymbol{\delta}_{s+1,c^\prime}^{\left(i+1\right)}\right)^T A_{\left(s+1,c\right),\left(s+1,c^\prime\right)}\boldsymbol{r}_{s+1,c}^{\left(i\right)}&=\Delta\varphi_{\left(s+1,c\right),\left(s+1,c^\prime\right)}^{\left(i\right)}\\
	\left(\boldsymbol{\delta}_{s,c}^{\left(i+1\right)}\right)^T A_{\left(s,c\right),\left(s+1,c^\prime\right)}\boldsymbol{r}_{s+1,c^\prime}^{\left(i\right)}+\left(\boldsymbol{\delta}_{s+1,c^\prime}^{\left(i+1\right)}\right)^T A_{\left(s,c\right),\left(s+1,c^\prime\right)}\boldsymbol{r}_{s,c}^{\left(i\right)}&=\Delta\varphi_{\left(s,c\right),\left(s+1,c^\prime\right)}^{\left(i\right)}\\
	\left(\boldsymbol{\delta}_{s,c}^{\left(i+1\right)}\right)^T A_{\left(s,c\right),\left(s-1,c^\prime\right)}\boldsymbol{r}_{s-1,c^\prime}^{\left(i\right)}&=\Delta\varphi_{\left(s,c\right),\left(s-1,c^\prime\right)}^{\left(i\right)}\\
	\left(\boldsymbol{\delta}_{s+1,c}^{\left(i+1\right)}\right)^T A_{\left(s+1,c\right),\left(s+2,c^\prime\right)}\boldsymbol{r}_{s+2,c^\prime}^{\left(i\right)}&=\Delta\varphi_{\left(s+1,c\right),\left(s+2,c^\prime\right)}^{\left(i\right)}.\label{eqSupLinear}
\end{align} 
with $c,c^\prime=1,..,C$. The first (second) row accounts for inaccuracies of the target implemented on cell $s$ ($s+1$), the third row accounts for crosstalk between cells $s$ and $s+1$, and the forth (fifth) row accounts for crosstalk between cell $s$ ($s+1$) and cell $s-1$ ($s+2$). The choice to correct the drive of two adjacent cells simultaneously is since this construction allows crosstalk between cells $s$ and $s+1$ to be mitigated simultaneously by both $\boldsymbol{\delta}_{s,c}$ and $\boldsymbol{\delta}_{s+1,c^\prime}$, which has been found to be more efficient than the correction of the crosstalk with control over a single qubit drive, e.g. the terms in the fourth and fifth rows. 

Crucially all of these terms result in $C\left(C-1\right)+3C^2$ linear constraints, and are independent of the total number of segments. Moreover iterative optimizations can be performed by considering independent blocks of 4 adjacent cells, that can be efficiently parallelized and interlaced. We remark that an additional linear constraint can be added in order to minimize the amplitude of the $\boldsymbol{r}_n$'s \cite{shapira2023fast}.

We perform $f$ optimization iterations, until convergence of the solution, or until meeting a fidelity criteria. We use the resulting solutions and benchmark the expected fidelity of the corresponding entanglement operations. The resulting infidelity is evaluated, in leading order, as,
\begin{align}
	I	&=I_{\text{short}}+I_{\text{long}}+I_{\text{motion}}\\
	I_{\text{short}}	&=2\sum_{s=1}^{S}\sum_{c,c^{\prime}=1}^{C+B_{B}}\left(\left|\Delta\varphi_{\left(s,c\right),\left(s,c^{\prime}\right)}^{\left(f\right)}\right|^{2}+\left|\Delta\varphi_{\left(s,c\right),\left(s+1,c^{\prime}\right)}^{\left(f\right)}\right|^{2}+\left|\Delta\varphi_{\left(s,c\right),\left(s-1,c^{\prime}\right)}^{\left(f\right)}\right|^{2}\right)\\
	I_{\text{long}}	&=2\sum_{s=1}^{S}\sum_{s^{\prime}\neq s,s\pm1}^{S}\sum_{c,c^{\prime}=1}^{C+B_{B}}\left|\Delta\varphi_{\left(s,c\right),\left(s^{\prime},c^{\prime}\right)}^{\left(f\right)}\right|^{2}\\
	I_{\text{motion}}	&=\sum_{j=1}^{N}\sum_{n=1}^{N}\left|\alpha_{j}^{\left(n\right)}\right|,
\end{align}
with $I_\text{short}$ representing the residual unwanted coupling between qubits that is accounted for by our mitigation iterations, $I_\text{long}$ representing the residual unwanted coupling between qubits that is not accounted for by our mitigation, and $I_\text{motion}$ accounting for unwanted residual displacement of the $j=1,...,N$ motional modes. Here $\alpha_j^{\left(n\right)}$ is the displacement of mode $j$ due to ion $n$ and is thus linearly related to the $\boldsymbol{r}_n$s and can be easily evaluated \cite{shapira2020theory,shapira2023fast}. While we are not directly controlling and minimizing the latter term in our optimization iterations, the resilience to motional frequency errors of the typical cell, discussed above, ensures that it remains small. 

\subsection{Crosstalk and power analysis}

We analyze the dependence of the crosstalk, in its mode-structure dependent formulation, $\varepsilon_{\text{J},b}$ in Eq. \eqref{eqCrosstalkJ} of the main text, on the choice of the number of barrier ions per cell, $B_A$, and on the optical confining power used per barrier ion, $\omega_\text{o.t.p}/\nu$.

\begin{figure}[!hbtp]
	\caption{Crosstalk dependence on $\omega_\text{o.t.p}$ and $B$. Left: Mean crosstalk (colored regions) for varying number of barrier ions (horizontal axis) and varying optical confinement, $\omega_\text{o.t.p}$, per barrier ions, given in units of the RF radial trapping, $\omega_\text{rad}$ (vertical axis). We observe that a perturbative crosstalk, set for values below 0.1 (area above white dashed line), is generated by setting $\omega_\text{o.t.p}\geq2\nu$, almost regardless of $B$. Nevertheless increasing $B$ is helpful in reducing the crosstalk. The total optical power per cell is proportional to $P=B\omega^2_\text{o.t.p}/\nu^2$ (colored lines, with the values of $P$ in the legend). Right: Estimation of band-dependent crosstalk, $\varepsilon_{J,b}$, based on mode structure analysis in an ion-crystal with cell size of $C=32$ qubits and $S=4$ segments, and a varying number of barrier ions, $B_A=B_B$ (color). In general the addition of barrier ions decreases the overall crosstalk in the system. Limiting cases are found for $B_A=1,2$ in which some of the collective modes do not involve the barrier ions, resulting in a non-typical crosstalk structure. Clearly for $B_A\geq3$ this behavior is suppressed. The mean crosstalk of the curves with of $B=2,4,6,8,10$ at $\omega_\text{o.t.p}\approx2.1\nu$ are correspondingly shown in the left.}
	\centering
	\includegraphics[width=0.75\linewidth]{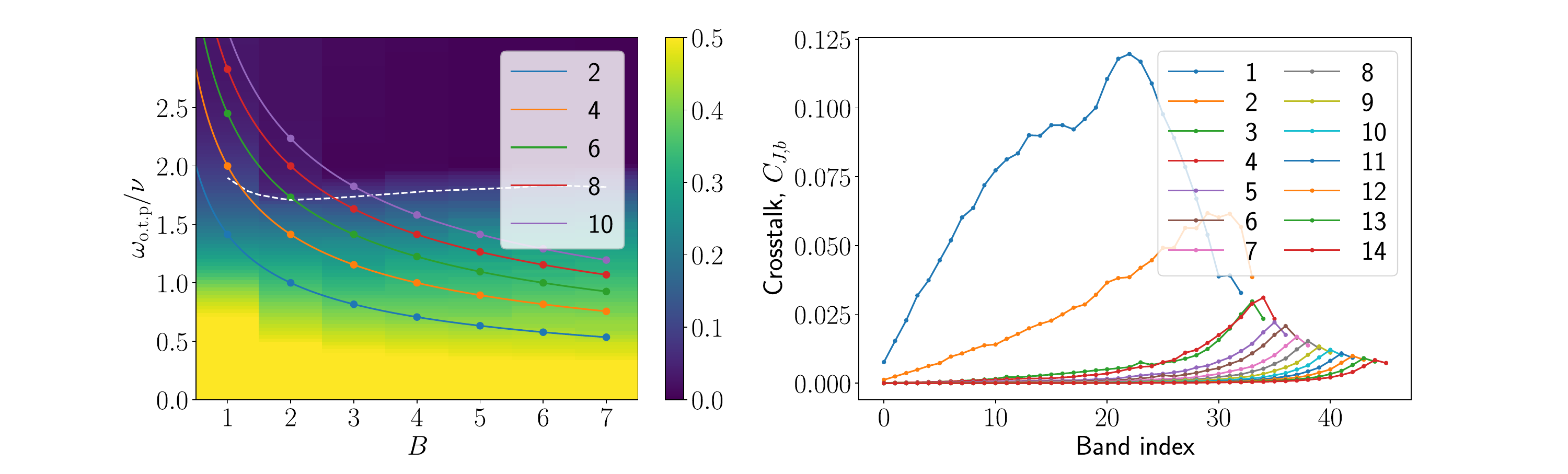}
	\label{figSupCrosstalkAnalysis}
\end{figure}

Figure \ref{figSupCrosstalkAnalysis} (left) shows the mean crosstalk averaged over all bulk-bands, $\langle \varepsilon_{\text{J},b}\rangle_b$ (color), for a segmented ion-crystal made of $S=4$ segments, each containing $C=32$ computational qubits and $B$ (horizontal axis) barrier ions, such that each barrier ion is illuminated by an optical confining potential with a trapping frequency $\omega_\text{o.t.p}$ (vertical axis). Clearly, as the optical trapping frequency decreases the mean crosstalk increases, and passes the perturbative limit, set here to $0.1$ (white dashed). We also present constant lines of the expression, $P=B\omega_\text{o.t.p}^2/\nu^2$, as it is proportional to the total required optical power per-cell (colored lines, see value of $P$ in the legend). 

Figure \ref{figSupCrosstalkAnalysis} (right) shows $\varepsilon_{J,b}$ for $C=32$, $S=4$ and various values of $B_A$, between 1 and 14 (color), such that for all $B_A$ barrier ions $\omega_\text{o.t.p}\approx2.1\nu$. The curves show expected crosstalk due to $b=1,..,C+B_B$ bands, with $B_B=B_A$, similarly to the inset of Fig. \ref{figRadialSpectrum}. The high-index cell bands are typically the main contributors to the crosstalk as they correspond to long-wavelength excitations of the cells, analogues to low-order oscillating multipoles, thus having a stronger coupling to adjacent cells. Clearly, a larger barrier reduces the overall crosstalk. Nevertheless, the curves do not show a monotonic behavior, which is especially obvious with $B_A=1$ (blue) and $B_A=2$ (orange), which exhibit resonant-like features. These resonances are likely due to the modes of motion in which the barrier-ions only weakly participate and therefore cannot isolate the motion of ions within a single cell. For $B_A\geq3$ this effect is generally suppressed, leading to only a residual distance-dependent Coulomb interaction between the cells. Furthermore, since our crosstalk mitigation technique relies on linearization, we have to work in the perturbative crosstalk regime, which is already shown to apply at $B_A\geq2$, motivating our choices in the main text.

\begin{figure}[!hbtp]
	\caption{Bandwidth and mode structure of the transverse mode of an optically segmented ion-crystal with $S=5$, $C=32$ and $B_\text{A}=B_\text{B}=3$. Left: Bandwidths of the $C=32$ bulk modes as a function of optical confinement. The bandwidths display a non-monotonic behavior stemming from a decrease of the bandwidth of a band as a barrier band crosses it. The inset shows a zoom-in on the bandwidth of the 4 top bulk band (vertical in log-scale) showing the global minimum of the bandwidth at $\omega_\text{o.t.p}=2\nu$. Right: Mode structure of the same system. Indeed the non-monotonic behavior of the bandwidths (left) are in correspondence with barrier bands crossing bulk bands. The inset shows a zoom-in on the vicinity of the minimal bandwidth of the center-of-mass band, showing indeed that his minimum is formed after crossing with the bulk band.}
	\centering
	\includegraphics[width=0.9\linewidth]{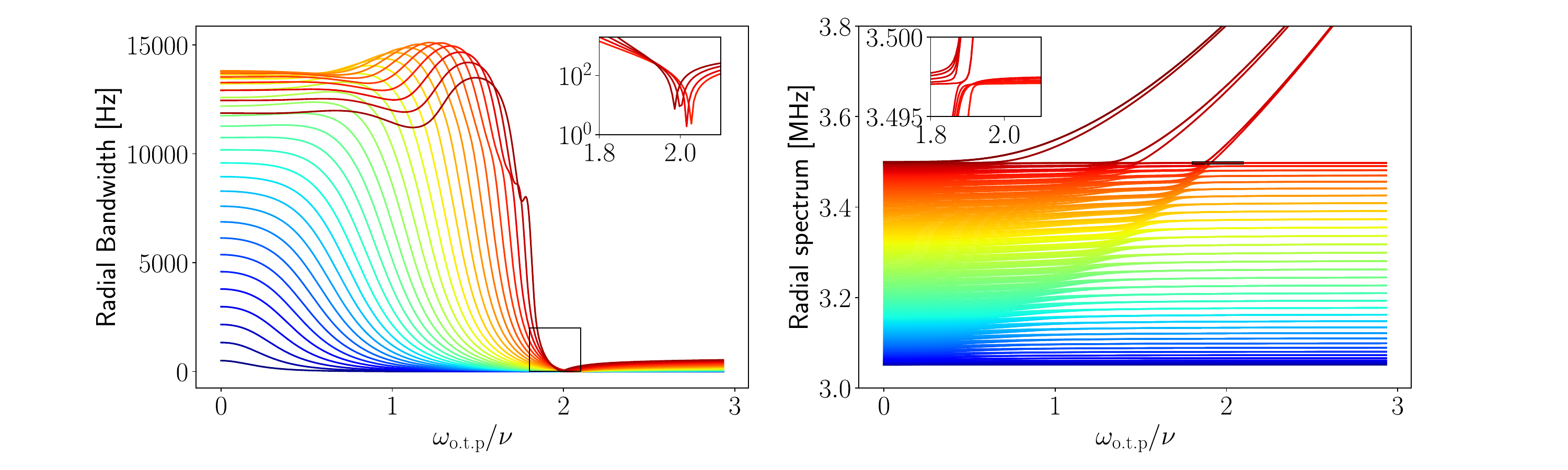}
	\label{figBandwidth}
\end{figure}

As shown in Fig. \ref{figMitigationAnalysis} of the main text, the optimal gate performance occurs at $\omega_\text{o.t.p}=2\nu$, and not at arbitrarily large optical trapping frequencies. This counter-intuitive effect occurs due to a narrowing of the bulk bands when they are close to the frequency of the barrier bands. Indeed. Fig. \ref{figBandwidth} (left) highlights this effect in the transverse modes of the system considered for the 3D Ising simulation, i.e. $S=5$, $B_\text{A}=B_\text{B}=3$ and $C=32$. The bandwidth (top) of the $C=32$ bulk bands (color), defined as the frequency difference between the highest and lowest modes in the band is shown as a function of optical trapping frequency. At the weak optical trapping regime, $0\leq\omega_\text{o.t.p}\leq2\nu$, increasing the optical frequency is shown to largely reduce the bandwidths, which directly contributes to a reduced crosstalk level between the segments. Already at this regime it is shown that the bandwidths have a non-monotonic behavior. The bandwidths reach a minima at $\omega_\text{o.t.p}=2\nu$, and then slightly increase. The inset shows a zoom-in on the bandwidths of the top 4 bands (vertical axis in log scale).

This behavior is explained by considering the frequency of the bulk and barrier modes, shown for the same system in Fig. \ref{figBandwidth} (right). Indeed, it is seen that the non-monotonic behavior of the bandwidth of a given band is exactly correlated with its crossing of a barrier band. The inset shows a zoom-in of the final barrier band crossing the top center-of-mass bulk band. Specifically it is seen that after the bands cross, at $\omega_\text{o.t.p}\approx1.9\nu$ the width of the band is minimal, and is slightly increased as optically confinement becomes stronger, leading to an increase of frequency of the barrier band.

\subsection{Reconfiguration rate of optical potentials}

Reconfiguration of the optical potentials may inadvertently excite the motion of computational qubit ions. We would like to determine the probability of exciting motion, and ensure that it is small. We note that this problem was already addressed in \cite{olsacher2020scalable} based on results in adiabatic theory \cite{adiabaticPT}; we therefore highlight only the main conclusion. 
In particular, the part of the ion-crystal Hamiltonian that is affected by the optical potentials is given by:

\begin{align}
H_\text{o.t.p}(t) = \sum_{i \in \{x,y,z\}} \sum_n \frac{1}{2}m\omega^2_\text{o.t.p}b^{(A)}_n r_{n,i}^2 + \frac{t}{\tau_R}\frac{1}{2}m\omega^2_\text{o.t.p}b^{(B)}_n r_{n,i}^2
\end{align}

Where $\tau_R$ is the reconfiguration time, $b_n^{(\{A,B\}))} = 1$ (0) if ion $n$ is (not) illuminated by an optical potential for each configuration, and the subscript $i$ denotes the trapping direction. Here we assume that optical potentials on the barrier ions of configuration B are ramped on while the barrier ions of configuration A remain illuminated (as prescribed by the mid-circuit measurement protocol). Furthermore, we assume that all motional modes of the ion crystal start in the ground state.
The probability of exciting $N > 0$ phonons in a given motional mode $m$ is given by \cite{olsacher2020scalable,adiabaticPT}:
\begin{align}
P^{(m)}_N \sim \hbar^2 \left( \frac {\bra{N(t)}\partial_t H_\text{o.t.p}\ket{0(t)}^2}{(E_{N,m}(t) - E_{0,m}(t))^4} \Bigg |_{t=0} + \frac {\bra{N(t)}\partial_t H_\text{o.t.p}\ket{0(t)}^2}{(E_{N,m}(t) - E_{0,m}(t))^4} \Bigg|_{t=\tau_R} \right )
\end{align}
Where $\ket{N(t)}$ are instantaneous eigenstates of the Hamiltonian with energy $E_{N,m}(t) = (N + \frac{1}{2})\hbar\omega_m(t)$, corresponding to $N$ excitations of motional mode $m$. We will simplify $\omega_m(t=0) \sim \omega_m(t=\tau_R) \equiv \omega_m$ as the mode frequency will not change drastically during the ramp. Focusing only on axial modes (as this will yield the highest probability of phonon excitation) we have:

\begin{align}
P^{(m)}_N \sim \frac{\hbar^2}{\tau_R^2} m^2 \frac{\omega_\text{o.t.p}^4}{(N \hbar \omega_m)^4} \left(  \bra{N(t = 0)}\sum_n r_{n,\text{axial}}^2 b^{(B)}_n \ket{0(t=0)}^2 + \bra{N(t = \tau_R)}\sum_n r_{n,\text{axial}}^2 b^{(B)}_n \ket{0(t = \tau_R)}^2 \right )
\end{align}

At any given time, in the instantaneous eigenbasis, the local position operator can be expanded as $r_{n,axial} = \sum_m A_{n,m}\sqrt{\frac{\hbar}{2m\omega_m}}(a_m^\dagger + a_m)$, where $A_{n,m}$ is participation of ion $n$ in axial mode $m$. The dominant contribution to the phonon excitation probability comes from the excitation of $N=1$ phonon:

\begin{align}
P^{(m)}_{N=1} \sim \frac{1}{\tau_R^2} \frac{\omega_\text{o.t.p}^4}{\omega_m^6}\left( (\sum_n A_{n,m}(t = 0)^2 b^{(B)}_n)^2 + (\sum_n A_{n,m}(t = \tau_R)^2 b^{(B)}_n)^2 \right ) \sim \frac{1}{\tau_R^2} \frac{\omega_\text{o.t.p}^4}{\omega_m^6} \left( \frac{B_B}{C} \right)^2 
\end{align}

For bulk modes the second sum will be small, as at time $t = \tau_R$ the barrier ions of configuration B will be illumminated and thus have a weak participation in the mode. The first sum conveys the participation of these ions while they are still non illuminated; generally $\sum_n A_{n,m}(t = 0)^2 b_n^{(B)} \sim \frac{B_B}{C}$. For typical configurations, $\frac{B_B}{C} \sim \frac{1}{10}$. In our case, the lowest axial frequencies correspond to $\omega_\text{c.o.m} \approx 200$kHz while $\omega_\text{o.t.p} \approx 1.5$MHz; thus $P^{\text{c.o.m}}_{N=1} < \frac{10^-9}{\tau_R^2}$. We conclude that using a reconfiguration time of $\tau_R = 100\mu$s is sufficient to ensure a low probability of exciting undesired motion. \par 
Even faster reconfiguration times can be used if employing a non-constant ramping profile. For example, we may consider a piece-wise linear profile consisting of two parts: in the first part, the optical potential is ramped (relatively slowly) to some intermediate value $\omega_\text{int}$ (where ideally the ion participation in bulk modes is already weak) at a time $\tau_\text{int}$, then in the second part it is ramped to $\omega_\text{o.t.p}$ in time $\tau_R-\tau_\text{int}$. Choosing $\omega_\text{int}$ and $\tau_\text{int}$ appropriately gives an order of magnitude improvement in the excitation probability and would therefore enable reconfiguration times of $\tau_R \sim 30 \mu$s. Certainly optimizing over different ramping profiles may yield further improvement.

\subsection{Example circuit breakdown}

Here we illustrate an example circuit within our architecture; we emphasize relevant heating mechanisms and cooling methodologies, and provide an estimate of the duration of each step within the circuit. Specifically, we focus on the QEC circuit mentioned in the main text and pictured in Fig. \ref{figQEC}. We highlight this specific example as it involves many of the important features of our architecture: high-connectivity multi-qubit entanglement gates, reconfiguration of optical potentials, and mid-circuit measurements. A generic mode of operation within this framework is that in configuration ‘A’ multi-qubit gates encode parity information of the code's plaquettes onto auxiliary ions, these ions are then optically confined and measured, implementing stabilizer measurements. Next the ion-crystal is reconfigured to configuration ‘B’, in which multi-qubit gates and measurements realize logical qubit-qubit interactions via lattice surgery, after which the ion-crystal is reconfigured back to configuration ‘A’. This process repeats until the desired algorithm is realized.
\par

There are several heating mechanisms that are relevant to this process, namely, heating of barrier ions due to off-resonance photon scattering induced by the optical trapping, leakage of this heat to the bulk modes, heating of the barrier ions due to on-resonance photon scattering when they are used as auxiliary ions and measured, and leakage of this heat to the bulk modes. Lastly, excitation of motional modes may occur during reconfiguration due to level crossing of the bulk modes, however this has already been treated in subsection E, above.\par

As mentioned in the main text, the optical potentials induce a $\Gamma_{\text{sc}}\sim3$kHz photon scattering rate on barrier ions. Each photon scattering event on ion $i$, assuming it is initially close to its ground state, has an $(\eta^{(i)}_m)^2 = (R^{(i)}_m)^2\eta_m^2$ probability of exciting a given motional mode $m$, where $\eta_m$ is its Lamb-Dicke parameter, and $R^{(i)}_m$ is the participation of the ion in the mode. The heating rate of a given mode due to this process is given by $\dot{n}_m = \sum_i \Gamma_\text{sc} (\eta^{(i)}_m)^2$. \par

The participation of barrier ions in bulk modes is very weak; accordingly, the average heating rate of bulk modes (taken over all $m$) due to $\Gamma_\text{sc}$ is approximately proportional to the number of barrier ions, yet small, $\braket{\dot{n}}\sim10^{-2}$ quanta per second.\par

Specifically, here we assume that multi-qubit gates are driven at the minimum possible gate time $T_{\text{min}} = 3.1$ms (where $T_{\text{min}}$ is inversely proportional to the minimum spacing of bulk bands \cite{shapira2023fast}). Each entanglement process within the QEC scheme (X stabilizers, Z stabilizers, lattice surgery) can be implemented using a single multi-qubit gate; these operations are implemented in parallel for all segments within the ion crystal. 
This implies an accumulated $\dot{n}_\text{Bulk}T_{\text{min}} \approx 3\cdot10^{-5}$ average quanta of motion during the gate, yielding a negligible impact on gate fidelity. \par

The barrier modes themselves, however, do heat as a result of this process and acquire an average of $\dot{n}_\text{Barrier}T_{\text{min}}\approx0.07$ quanta over the gate duration; these modes should therefore be cooled before the optical potentials are reconfigured and the barrier ions join the bulk. \par

In the QEC scheme above there are $12$ auxiliary ions per segment that must be measured after each stabilizer operation. The mid-circuit measurement procedure begins by illuminating each of these ions with an optical trapping beam. We assume that measurements are carried out by exposure to $397$ nm light, where the collection of $\sim 30$ photons per ion (e.g. using a multi-channel photo-multiplier tube) is sufficient to determine that the ion is in the "bright" state. Further assuming a photon collection efficiency of $1 \%$, this corresponds to $3000$ scattered photons per measured ion.\par 

In order to mitigate anti-trapping effect in the $P$ levels, we set the scattering rate to be low enough, such that the ions spend only a small fraction of the detection time in the $P$ levels. Specifically, by spreading the detection duration over $1$ ms, the ion will occupy the $P$ level for $\frac{3000\tau}{1 \text{ms}} = 2.4\%$ of the detection time, where $\tau \approx 8$ ns is the lifetime of the $P$ level. Since the anti-trapping forces in the $P$ level are only half as strong as trapping forces in the $S$ levels, this implies that the reduction of the effective optical trapping potential during detection will be at the $5\%$ level. \par

There are two sources of heating associated with mid-circuit measurement. First, each of the $12$ measured ions may scatter $3000$ photons; since measurement is performed with red-detuned light which also provides Doppler cooling, the ions would only heat up to the Doppler temperature corresponding to $\bar{n}_D = 5$ quanta. We note that this is the "worst case" in terms of heating. In general, we are free to choose stabilizer eigenvalues such that all measured ions are dark except in the rare case of a detected error; in this configuration, photon scattering will hardly occur. As the measured ions are optically confined, this heat remains local, and minorly affects the modes of computational qubit ions. In particular, assuming each measured ion is locally at the Doppler temperature, we estimate the heat accumulated on a mode $m$ of the computational qubit ions to be $\bar{n}_m = \sum_i (R^{(i)}_m)^2 \bar{n}_D$. On average the modes of computational qubit ions will heat by $\braket{\bar{n}} = 0.005$ quanta due to the mid-circuit measurements. This heating is negligible, yet may accumulate over many rounds of operation, and is thus mitigated below.\par

Another source of heating during mid-circuit measurement is related to the fluctuating dipole force on the measured ion \cite{Hutzler_2017,Alt_thesis}. Each time the ion is excited to the $P$ level, assuming it is slightly off-center of the optical trapping beam, it experiences a change in the dipole force $\Delta F \approx m\omega^2_\text{o.t.p}x$ (where $x$ is the ion displacement from the center of the beam). This change of force occurs over the lifetime of the $P$ levels and translates to a momentum kick $p_\text{dip} = \Delta F \tau$. Taking the displacement to be $x=0.25\mu $m (an extreme limit corresponding to the steepest slope of the Gaussian beam) and comparing to the photon recoil momentum $p_\text{rec} = \hbar k_{397}$, we have $p_\text{dip} \approx .15*p_\text{rec}$. We notice that even for a large displacement, this effect is smaller than photon recoil. \par

As optically confined ions (both barrier and measured ions) may heat considerably, they must be cooled during the circuit. Ground state cooling the modes of all optically confined ions can be done in parallel within the ‘prepare’ step following mid-circuit measurement. Based on efficient cooling techniques capable of cooling multiple motional modes simultaneously \cite{Wu2023cooling,Che2017cooling,Rasmusson2021cooling}, we estimate this will require $\sim 500 \mu$s. Reconfiguration of the optical potential is performed in two stages of $	\gtrsim 100\mu$s (as described in the previous section). An additional benefit of this protocol is that when the optical potential is removed from the cooled ions, they will in turn cool other modes of the ion crystal sympathetically, thus mitigating the small heating effects sustained during gate duration and mid-circuit measurements. \par

We estimate that the temperature of computational qubit ions will reduced by the ratio of number of cooled ions to total number of ions in the crystal, which in the case of this circuit is $\frac{14}{39} \approx 36\%$. In all, heating of computational qubit ions during the gate and mid-circuit measurement is at the $10^{-3}$ quanta level. This would permit 100s of rounds of mid-circuit measurements before having an appreciable effect on gate fidelity and necessitating cooling of the full ion crystal. In practice, the physical heating rate of any real ion trap device provides a similar or stricter limitation. \par

Focusing on operation times, a round of QEC stabilizers – including mid-circuit measurement, state preparation, ground state cooling, and optical potential reconfiguration – would require approximately $5$ms. Incorporating an additional layer to implement logical qubit entangling gates (e.g. via lattice surgery) would bring the total circuit time to $10$ms. Here the majority of the circuit is taken up by the entanglement gates. This should be contrasted to QCCD architectures, where the gate times can be small ($\sim10 \mu$s) however the majority of the circuit duration is dedicated to ion shuttling and cooling operations. For example, in Ref. \cite{moses2023race} a QCCD approach featuring two-ion trapping sites, reports that $97-99 \%$ of the circuit is dedicated to ion transport and cooling for a diverse range of circuit examples. Based on the results in that reference (particularly the circuits analyzed in Table I there), we can roughly estimate that the same QEC circuit considered here would require $100$'s ms using that approach. It should be noted that the exact duration of a shuttling based circuit highly depends on the ordering of shuttling operations and the corresponding optimization of this process. Moreover, it is possible to consider a shuttling-based approach featuring transport of long ion crystals as opposed to two-ion segments; this would reduce the number of required shuttling operations \cite{murali2020architecting}. However, the complexity of ion transport and the associated heating rate also increases with the segment size. \par

Another important comparison should be made to an architecture based on photonic interconnects. In this method, many trapping modules are linked via a photonic network - where each trapping module includes computational qubit ions, and ions dedicated to photon-mediated communication between modules. Based on Ref. \cite{monroe2014photons}, each remote entanglement operation between two communication qubits in different trapping modules would require $3$ms. Overall circuit times highly depend on the structure of the trapping modules. In particular, local entangling gate times depend on the size of each register; moreover implementing local and remote gates, as well as mid-circuit detection may require ion shuttling and cooling operations as in QCCD. Nonetheless we can roughly compare to a similar setup, with a single logical qubit contained in each local trapping module. We assume remote entanglement is done in parallel; this would require 10 communication ports and 10 additional communication qubits per module. (We note that the hardware overhead can be mitigated at the cost of running remote gates sequentially, and thus having larger overall circuit duration). Due to the increased crystal size, the minimum local entangling gate time increases to $\sim 5$ ms. Lattice surgery then involves parallel remote entanglement in addition to a local entangling operation and measurement. Overall we estimate this circuit would run in $\sim 20$ ms. If remote entanglement is performed sequentially this time would increase to $\sim 100$'s ms. \par 

A comparison between our architecture, QCCD, and a photonic interconnect approach is summarized in Table \ref{tbl:comparison}. The total circuit time corresponds to the example QEC circuit detailed above. \par

\begin{table}[!hbtp]
  \caption{Comparison between our architecture (optical segmentation), QCCD, and a photonic interconnect approach. We focus on the QEC example circuit (involving stabilizer measurements and lattice surgery) as a basis for the comparison. The resource estimates for the latter two architectures are derived from Refs \cite{moses2023race} and \cite{monroe2014photons} respectively. We emphasize that times shown here are a rough estimate and could vary based on the details of the quantum circuit. This comparison shows the large potential speedup offered by our method, which is due to the low overhead of connecting remote qubits.}
  \label{tbl:comparison}
  \includegraphics[width=0.95\linewidth]{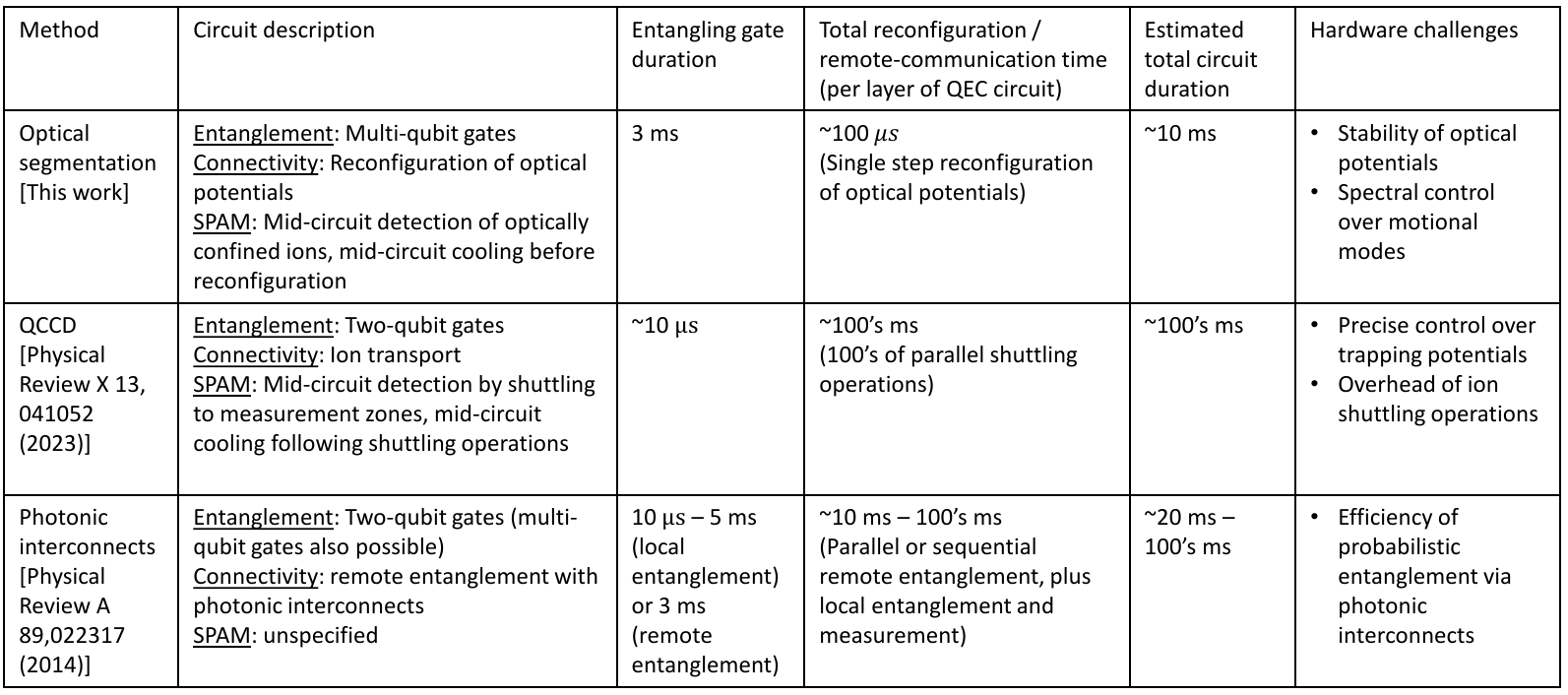}
\end{table}

While only intended as an approximate resource estimate, this example circuit highlights the feasibility of our architecture under realistic experimental conditions, and a potential advantage over other methods in terms of overall circuit duration.

\end{document}